\documentclass[11pt, a4paper]{article} 

\usepackage{fullpage}

\usepackage[onehalfspacing]{setspace}

\usepackage{palatino}

\usepackage{authblk}

\usepackage{amsmath,amssymb,amsfonts,amsthm,makeidx}
\usepackage{txfonts}
\usepackage{tikz-cd}
\usepackage{braket}
\usepackage{bm}
\usepackage{hyperref}
\usepackage{color}
\usepackage{xcolor}
\usepackage{comment}
\usepackage{graphicx}
\usepackage[export]{adjustbox}
\usepackage{booktabs, array}
\usepackage{enumerate}
\usepackage{subcaption}
\usepackage[shortlabels]{enumitem}

\begin{document}

\title{Scalar Quantum Fields: Theory Space and its Geometry}

\author[1]{Viola Gattus}%
\author[2]{Peter Millington}%

\affil[1]{Technische Universit\"at M\"unchen, School of Natural Sciences, Garching 85748, Germany}
\affil[2]{University of Manchester, Department of Physics and Astronomy, Manchester M13 9PL, United Kingdom}

\maketitle
 
\begin{abstract}

Scalar fields provide perhaps the simplest playground in which to develop our understanding of quantum field theory. In this lecture, we consider what it means to write down a scalar quantum field theory and how we can give geometrical interpretations to the space of such theories:\ the theory space.

\end{abstract}

\vspace{0.2em}
\noindent\rule{\linewidth}{0.3pt}
\vspace{0.01em}

{\scriptsize
\noindent
This collection of materials has been prepared by the authors as an initial draft for the forthcoming combined volume ``Comprehensive Quantum Physics,'' to be published by Elsevier Ltd. \par}
\vspace{0.05em}
\noindent\rule{\linewidth}{0.3pt}
\vspace{0.2em}

\section{Introduction: Theory space}

How do we build theories? As the whimsical but nevertheless wise Winnie-the-Pooh might say, before one can come to a theory, one might first like to know what a theory is and whether there is any reason to fear such a thing, supposing, as a matter of fact, a theory were ever to jump out from behind a gorse bush and say, ``How do you do?.'' 
Rest assured, the theories we consider in this lecture are all of upright character, but defining what we mean by a theory is not a bad place at all for us to start. So let's start there.

Often, in the spirit of logical positivists like Ernst Mach, we talk of wanting to explain empirical observations. For those of us engaging with this material, the observed phenomena might be the properties and interactions of elementary particles. But actually, we might want to do more than explain \emph{observed} phenomena; we might want to predict \emph{new} phenomena for which we (or at least those of us with a flair for experiment or observation) can go off and look. The history of particle physics has seen theories play both these roles \cite{Frampton:2020kki}.

So, for us, a theory will be a mathematical description of the dynamics of elementary particles that takes some inputs, which we might call model parameters, applies some rules, which we will give some fancy title like \emph{relativistic quantum mechanics} or \emph{quantum field theory}, and makes some prediction. So we have three things to do: (i) set down the mathematical language and rules, (ii) decide on the input parameters (i.e., build the model) and (iii) derive the relationships between those input parameters and physical observables --- (iii) is not independent of (i).

Now, there was a time, not so long ago, when the prevailing attitude was that the only good theories were ones that had a finite number of input parameters and a much larger number of physical observables. This is a win for predictivity, since we can eliminate the input parameters, stick with Mach, and concentrate only on the relationships between physical observables. Theories like this are highly coveted within the reductionist philosophy that has shaped much of modern physics, and they fall into two categories, each given the less-than-transparent labels of \emph{perturbatively renormalizable} or \emph{non-perturbatively renormalizable}. %(You may want to check out the companion chapters to learn more about these.) 
The flagship example of a perturbatively renormalizable theory is quantum chromodynamics (QCD), which describes the strong interaction. This theory is one of the lucky ones:\ it is \textit{asymptotically free}, which implies that at very high energies --- what one would call the \textit{ultraviolet (or UV) regime} --- the strong coupling that controls the strength of the interactions becomes progressively smaller, quarks and gluons behave as weakly interacting particles, and we are free to use all the tricks from perturbation theory. The second category includes those theories that are a little less “standard” \cite{Mastropietro:2008}. They possess the property of \emph{asymptotic safety}, which guarantees the existence of a finite number of so-called relevant directions and, in turn, a finite number of input parameters. A textbook example of a perturbatively non-renormalizable but non-perturbatively renormalizable quantum field theory is the three-dimensional Gross--Neveu model \cite{Braun:2010tt}. But do not be pushed away by the tongue-twister:\ such theories are of unarguable interest to study.

But attitudes inevitably change --- sometimes for the better and sometimes for the worse --- and it turns out that there are many physically relevant theories that do not find themselves lucky enough to be part of these cliques. These theories are called \emph{effective}, and they are in good company, alongside General Relativity (still our best description of classical gravitation), chiral perturbation theory (describing the dynamics of hadrons), the Ginzburg--Landau theory of superconductivity (we have plenty to learn from our condensed-matter colleagues) and, indeed, the much-celebrated Standard Model of particle physics if we consider it the low-energy limit of some larger theory (that might, for example, offer an explanation of the origins of dark matter or dark energy). The latter is captured in the Standard Model Effective Field Theory \cite{Brivio:2017vri}. Quantum electrodynamics (QED) would also fall into the category of effective theories in the modern sense. While QED is perturbatively renormalizable, there is a catch:\ in perturbation theory, it develops a \textit{Landau pole}, and the QED fine-structure constant mathematically diverges at finite, albeit arbitrarily high, energies \cite{Peskin:1995ev}. The presence of this Landau pole suggests that QED cannot be regarded as a UV-complete theory on its own. 

Wound up in understanding these classifications of theories is the appreciation that the parameters of our theories and their relationships to physical observables depend on the characteristic energy scale of our measurements:~this is at the heart of \emph{renormalization} (we'll come back to this later).

Let's turn then to our little excursion into \emph{quantum field theory}. What is it that we want our theories to describe? Ultimately, we want to write down theories that describe the microscopic origins of the known fundamental forces. We'll leave out gravity %--- that would need another book of its own (or at least the chapter that appears later in this collection) --- 
and we'll stick to toy models, too, and just \emph{scalar} fields (describing particles with zero spin), since higher spins are a technical complication to which we do not need to subject ourselves just yet. This will still be more than enough for us to gain some understanding of what we mean by \emph{theory space}.

Now to complete step (i):\ set down the mathematical language and rules. When we said earlier that we would be writing down theories of the dynamics of the elementary particles, we weren't being entirely honest. In fact, we will write down theories of fields --- whatever they are --- and we will try to interpret the excitations of these fields as elementary particles.

There are a number of ways that we can approach quantum field theory:
\vspace{-0.5em}
\begin{enumerate}[(i)]
    \item the canonical operator formulation;
    \vspace{-0.5em}
    \item the path integral formulation;
    \vspace{-0.5em}
    \item the operator product expansion.
    \vspace{-0.5em}
\end{enumerate}
We will make use of the first two, and leave the third to people who know about conformal field theory.

So, how do we write down a quantum field theory? We would be forgiven for thinking that we just need to write down an action, say
\begin{equation}\label{eq:action}
    S[\Phi]\,=\int{\rm d}^4x\left[\eta^{\mu\nu}\,Z_{ab}\,\partial_{\mu}\Phi^a\, \partial_{\nu}\Phi^b-V^{(1)}_a\,\Phi^a-V^{(2)}_{ab}\,\Phi^a\,\Phi^b-V_{abc}^{(3)}\,\Phi^a\,\Phi^b\,\Phi^c-V^{(4)}_{abcd}\,\Phi^a\,\Phi^b\,\Phi^c\,\Phi^d\,+\,\ldots\right]\,,
\end{equation}
where $Z_{ab}$ is some kinetic mixing matrix and the $V$'s label the interactions.  
Here, the square brackets indicate that the action is a \emph{functional} of the field $\Phi\equiv \Phi(x)$, i.e., it maps functions of the spacetime coordinate $x^{\mu}$ to numbers, $\eta^{\mu\nu}={\rm diag}(1,-1,-1,-1)$ is the Minkowski metric, and we consider $N$ real scalar fields $\Phi^a$ with $a\in\{1,2,\dots,N\}$. We have certainly written down a field theory, but we have not yet written down a \emph{quantum} field theory. The latter is composed of two things:
\vspace{-0.4em}
\begin{enumerate}[(i)]
    \item some generators of the dynamics (in Minkowski spacetime, these are the generators of the Poincaré group and, in particular, the Hamiltonian, the generator of time translation, for which the action will serve as a proxy);
\vspace{-0.4em}
    \item some state space around which the generators of dynamics move us (this will be something like a Hilbert space, but there are some subtleties related to how we deal with continuous spectra, for example, that we won't go into in this lecture).
    \vspace{-0.4em}
\end{enumerate}
We often forget that there is freedom in how we choose \emph{both} of these ingredients of our quantum field theory, and the challenge is to find useful bases to span the state space. (We will consider one choice below.) However we choose these ingredients, we nevertheless have our first way of charting our \emph{theory space}:\ we can use the parameters of the theory as coordinates, or more precisely the coefficients of the various ``operators'' appearing in the action. These are just the $Z_{ab}$ and $V$'s in \eqref{eq:action}.

One choice for spanning the state space is the basis of \emph{field eigenstates}, which is useful for constructing the path integral, which we will come to shortly. Field eigenstates satisfy the eigenequation (taking $N=1$ for now)
\begin{equation}
    \hat{\Phi}(t,\mathbf{x})\,\ket{\,\Phi(\mathbf{x}),\,t}\,=\,\Phi(\mathbf{x})\,\ket{\,\Phi(\mathbf{x}),t}\,.
\end{equation}
The state $\ket{\,\Phi(\mathbf{x}),t}$ is an instantaneous eigenstate of the Heisenberg-picture field operator $\hat{\Phi}(t,\mathbf{x})$ at time $t$ with ``eigenvalue'' $\Phi(\mathbf{x})$. In accordance with the Heisenberg picture, the operators carry all the time dependence, while the states remain constant in time. Therefore, the $t$ in $\ket{\,\Phi(\mathbf{x}),t}$ should not be taken to indicate dynamical time evolution of the eigenstate. Note that the field eigenstate is also not really a function of the spatial coordinate $\mathbf{x}$, but rather a vector in the Hilbert space labeled by the entire field configuration $\Phi(\boldsymbol{\mathrm{x}})$. We include the argument to remind us that the eigenstate encodes information about the configuration of the system at all points in space.

The field eigenstates are orthonormal and complete in a functional sense, with
\begin{equation}
    \int \left[{\rm d}\Phi_t(\mathbf{x})\right]\,\ket{\,\Phi(\mathbf{x}),t}\bra{\Phi(\mathbf{x}),t\,}\,=\,\mathbb{I}\,,
\end{equation}
where $\left[{\rm d}\Phi_t(\mathbf{x})\right]\,=\,\prod_{\mathbf{x}}{\rm d}\Phi_t(\mathbf{x}) $ is a formal functional integral measure (with the subscript $t$ understood as a label) and $\mathbb{I}$ is the infinite-dimensional unit operator on the Hilbert space of the field theory. We might be tempted (and, in fact, we should be) to ask why we tend to model build only at the level of the classical action and not also in terms of the way we build the state space. In fact, more generally, we can introduce a metric on the state space, which will play a role similar to a weight function in the orthonormality of the basis. This is the approach of so-called pseudo-Hermitian and PT-symmetric quantum theories~\cite{Mostafazadeh:2001jk, Bender:2005tb}, which have found a wide range of applications~\cite{El-Ganainy:2018ksn, Ashida:2020dkc}. These theories are constructed in terms of general antilinear symmetries,\footnote{An antilinear transformation $K$ complex conjugates complex numbers, i.e., for all complex $\lambda$, $K\lambda K^{-1}=\lambda^*$.} such as the combined action of parity (P) and time-reversal (T), which replace the mathematical condition of Hermiticity.\footnote{As an example, a Hermitian matrix $A$ is equal to the transpose of its complex conjugate:\ $A=(A^*)^{\mathsf{T}}\equiv A^{\dag}$.}

\begin{figure}
\centering
\includegraphics[width=0.45\textwidth]{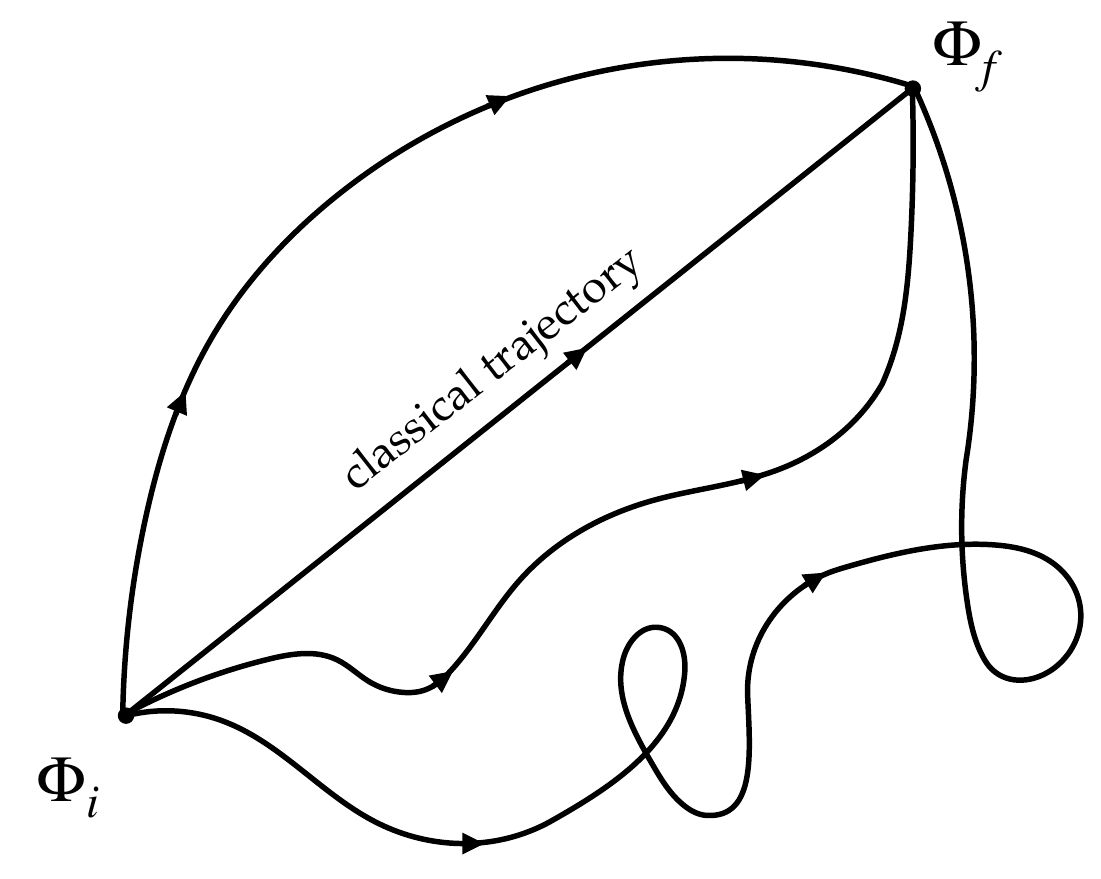}
\caption{An example of possible paths between initial and final field configurations $\Phi_i$ and $\Phi_f$. Classically, the field follows a single trajectory: the shortest path. In the quantum regime, however, an infinite number of paths contribute. The sum over all such possible paths is known as the path integral.}\label{fig:paths}
\end{figure}

Returning to our field eigenstates, and in analogy to quantum mechanics, one might wish to compute the evolution of an initial field eigenstate $\Phi_i$ into another field eigenstate $\Phi_f$ under the action of some Hamiltonian operator $\hat{H}$. This \textit{transition amplitude} is given by the matrix elements
\begin{equation}
    \bra{\Phi_f(\mathbf{x})\,}\,\mathrm{e}^{-i \hat{H} t/\hbar} \,\ket{\,\Phi_i(\mathbf{x})} \,=\, \int \left[\mathrm{d}\Phi\right]\,\mathrm{e}^{i S[\Phi]/\hbar}\,,
\end{equation}
where $S[\Phi]$ was defined in \eqref{eq:action} and $[\mathrm{d}\Phi]=\prod_{t=t_i}^{t_f}[\mathrm{d}\Phi_t]$. (Herein, we have in mind that $t_i\to -\infty$ and $t_f\to\infty$.) The right-hand side of this expression is what we term the ``path integral.'' It owes its name to the fact that in quantum mechanics (and hence quantum field theory), there exist infinitely many ways or ``paths'' by which a system may evolve from one state to another, in contrast to the unique trajectory of classical mechanics. The path integral is then defined to be just a sum (in the continuous limit) over all the paths between the field eigenstates $\Phi_i$ and $\Phi_f$ with weight prescribed by the exponentiated action functional $\mathrm{e}^{i S[\Phi]/\hbar}$. Figure~\ref{fig:paths} provides a schematic illustration of this idea.

Now that we have introduced the concept of the path integral, let us continue in our aim of building a quantum field theory. Step (ii), building the model, amounts to deciding which operators to include and considering how we might be guided to choose them. However, we are also interested in how the system evolves, and one way to find this out is to kick it and see what happens. The slightly more scientific approach to kicking a quantum field theory is to introduce external sources $J_a$, which perturb the system away from its ground state, and then measure changes in the system. One way to do this is to consider correlations in the system, and a convenient way to generate expressions for these \emph{correlation functions} is to introduce the particular path integral
\begin{equation}\label{eq:1p1_func}
Z[J]\,=\,\mathcal{N}\int\left[\mathrm{d}\Phi\right]\,\exp\left[\frac{i}{\hbar}\left(S[\Phi]-\int{\rm d}^4x\, J_{a}(x)\,\Phi^{a}(x)\right)\right]\,,
\end{equation}
where $\mathcal{N}$ is some suitable normalization factor. We can then obtain the correlation functions by taking functional derivatives with respect to the sources. But let's not go too fast. The functional integral over field configurations $\Phi$ is not convergent, because the integrand is oscillatory. We can tame this by taking $V^{(2)}_{ab}\to V^{(2)}_{ab}-\delta_{ab}\,i\epsilon/2$, where $\epsilon\,=\,0^+$. This is the Feynman prescription, and it will have two effects:\ first, it will allow us to invert the Klein--Gordon operator
\begin{equation}
    \frac{\delta^2 S[\Phi]}{\delta \Phi^{a}(x)\,\delta \Phi^{b}(y)}
\end{equation}
to obtain the tree-level two-point correlation function; and second, it will give us time ordering of the resulting correlation functions. Moreover, this prescription allows us to construct the Wick contour to rotate to Euclidean space by analytically continuing $t\to -i\tau$ (the process is actually much more subtle than this), with $\tau$ being Euclidean time. We can then identify the classical Euclidean action with the classical free energy, and it is for this reason that we often refer to $Z[J]$ as the \emph{partition function}. The analogy with statistical mechanics becomes exact when we interpret the Euclidean time direction as inverse temperature. We will not do that here.

With this prescription, $Z[J]$ acts as a generating functional of time-ordered correlation functions. More specifically, we can obtain the $n$-point time-ordered correlation function via
\begin{equation}
    \braket{\mathrm{T}[\Phi^{a_1}(x_1)\,\Phi^{a_2}(x_2)\ldots \Phi^{a_n}(x_n)]}_J\,=\,\frac{1}{Z[J]}\left(\prod_{j\,=\,1}^n i\hbar\frac{\delta}{\delta J_{a_j}(x_j)}\right)\,Z[J]\,,
\end{equation}
where the operator $\mathrm{T}$ ensures that the fields are ordered with the latest time to the left.
However, the correlation functions obtained from $Z[J]$ will also contain disconnected pieces. To generate only connected correlation functions, we can vary the \emph{Schwinger functional}
\begin{equation}\label{eq:1p1_schwinger}
    W[J]\,=\,i\hbar\ln Z[J]\,,
\end{equation}
whose Euclidean counterpart is related to the free energy. For example, the connected two-point function is
\begin{align}\label{eq:2pt_1pi}
    \hbar \Delta^{ab}(x,y)\,&=\,i\hbar\,\frac{\delta^2W[J]}{\delta J_{a}(x)\,\delta J_{b}(y)}\,=\,\frac{(i\hbar)^2}{Z[J]}\,\frac{\delta^2Z[J]}{\delta J_{a}(x)\,\delta J_{b}(y)}\,-\,\frac{(i\hbar)^2}{Z^2[J]}\,\frac{\delta Z[J]}{\delta J_{a}(x)}\,\frac{\delta Z[J]}{\delta J_{b}(y)}\nonumber\\
    \,&=\,\braket{\mathrm{T}[\Phi^{a}(x)\,\Phi^{b}(y)]}_J\,-\,\braket{\Phi^{a}(x)}_J\braket{\Phi^{b}(y)}_J\,.
\end{align}

Of course, we can add any sources we want, and it will prove useful to introduce a source for the two-point correlation function. To this end, we generalize the partition function to
\begin{equation} \label{eq:partition_jk}
     Z[J,K]\,=\,\int\left[\mathrm{d}\Phi\right]\,\exp\left[\frac{i}{\hbar}\left(S[\Phi]\,-\int{\rm d}^4x\, J_a(x)\,\Phi^a(x)\,-\,\frac{1}{2}\int{\rm d}^4x\,{\rm d}^4y\, \Phi^a(x)\, K_{ab}(x,y)\,\Phi^b(y)\right)\right]\,,
\end{equation}
with the Schwinger functional becoming
\begin{equation}
    W[J,K]\,=\,i\hbar \ln Z[J,K]\,.
\end{equation}
(We could keep going and add a three-point source and so on. Let's not.) We can now obtain the two-point function from a single variation with respect to the two-point source $K(x,y)$; namely,
\begin{equation}
    \braket{\mathrm{T}[\Phi^{a}(x)\,\Phi^{b}(y)]}_{J,K}\,=\,2\,\frac{\delta}{\delta K_{ab}(x,y)}\,W[J,K]\,=\,\frac{2i\hbar}{Z[J,K]}\, \frac{\delta}{\delta K_{ab}(x,y)}\,Z[J,K]\,.
\end{equation}

However we choose to obtain the correlation functions, we now have the second way of charting our theory space:\ we can use the correlation functions of the theory as coordinates. Moreover, in having written down the generating functionals in the path-integral representation, we are now well on our way to completing step (iii).

Before continuing, one might wonder whether there exists an intuitive, pictorial way of representing particle interactions and quantum corrections. To this end, recall that the potential $V$ appearing in the action functional \eqref{eq:action} encodes the interactions of the fields. Expanding the action in powers of the fields leads to interaction terms, which can be conveniently represented using \textit{Feynman diagrams}. Feynman diagrams provide a powerful visual and computational tool that serves as a bridge between the canonical operator formalism and the path integral formulation of quantum field theory. In these diagrams, fields are represented by lines (propagators), while interaction terms in the action correspond to vertices. External lines represent incoming or outgoing particle states, whereas internal lines represent the contributions of virtual particles that we must integrate over. Quantum corrections appear in the form of closed ``blobs'' known as \textit{loops} in these diagrams.  Such loops arise from integrating over internal degrees of freedom and encode the effects of quantum fluctuations around classical interactions, and we will see briefly in the next section how these structures arise in the case of a simple scalar field theory. 

\section{The quantum effective action}
As we anticipated in the previous section, the partition function defined in \eqref{eq:partition_jk} can be used to generate $n$-point correlation functions. However, it would be nice if, instead of having something that was a function of the sources, we had an object that was a function of the correlation functions themselves. It would be even better if this thing were akin to the classical action, whose variation gave us the equations of motion governing those correlation functions. We're in luck:\ this object exists, or rather a family of such objects exists, and we call them \emph{quantum effective actions} or just \emph{effective actions}.

\begin{figure}
\centering
\includegraphics[width=0.8\textwidth]{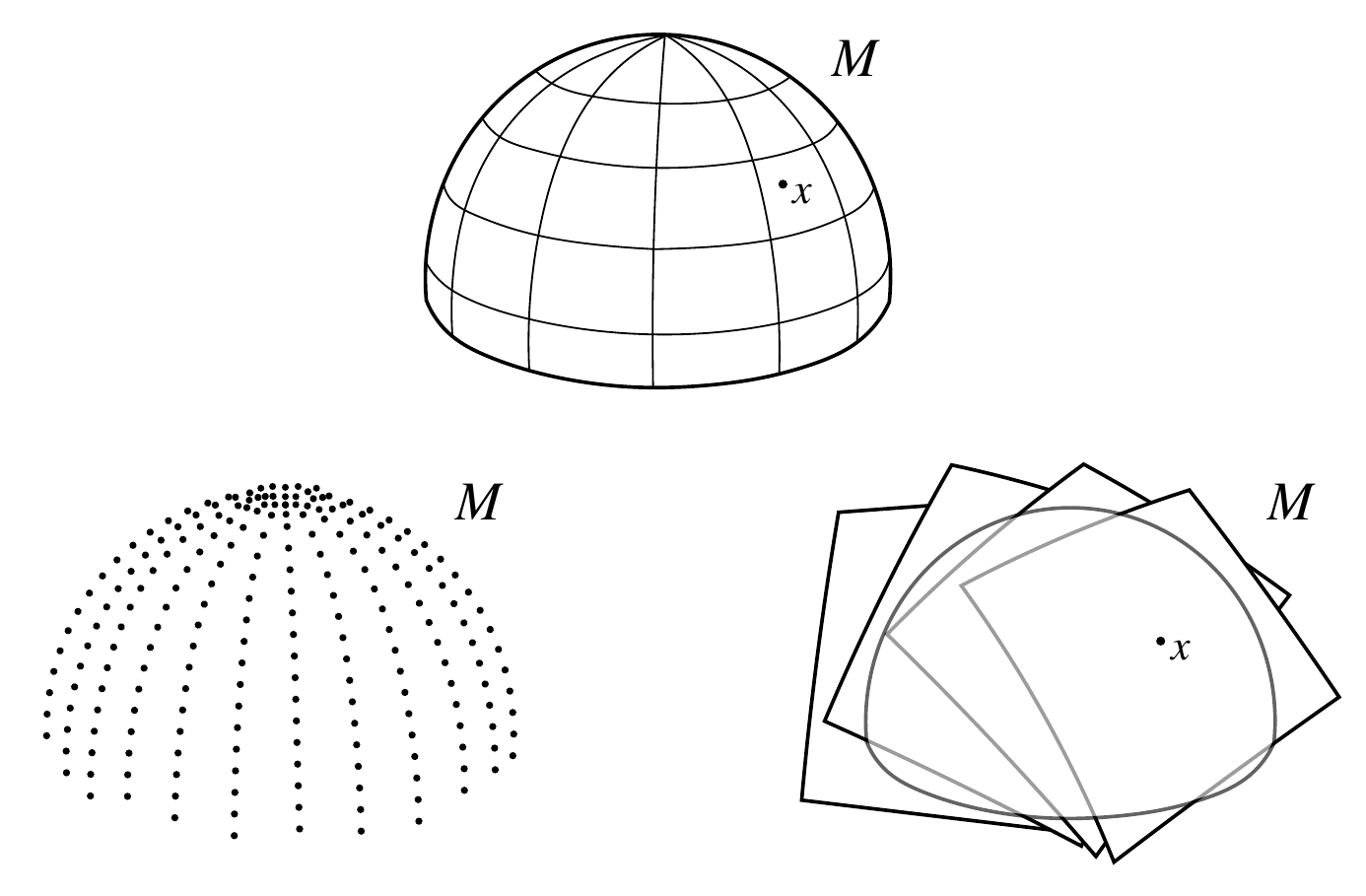}
\caption{Schematic depiction of two different ways to ``build'' a manifold $M$, represented here as a hemisphere. The manifold $M$ can be constructed either by specifying all its points $x$ or by drawing the tangent plane at each point $x$ on its surface.}\label{fig:sup_charts}
\end{figure}

To construct them, we first need to take a little diversion into differential geometry. Consider a point $x$ on some manifold $M$. We can define a tangent space at the point $x$. As an example, suppose the manifold $M$ is a hemisphere. The tangent space at a point $x$ on the hemisphere is just the plane tangential to the surface at that point. Now, we could build the hemisphere by giving the set of points $x$ or from the envelope of the set of all planes tangential to the surface. Figure \ref{fig:sup_charts} illustrates this scenario schematically. 

We can do the same thing with convex or concave functions:\ we can build a function $f(x)$ from the coordinates $(x,f(x))$ or we can build it from the set of tangents to the function, specified in terms of their gradients and intercepts. The mapping between these two sets of coordinates is a \emph{Legendre transform} \cite{Zia:2009}.

In the scenario at hand and for the applications of interest to us, the Schwinger functional $W[J,K]$ will play the role of the function $f(x)$. The tangents to $W[J,K]$ are found through the extremization 
\begin{align}
    \phi^a(x)\,&\equiv\, \langle \Phi^a(x)\rangle_{J,K} \,=\, \frac{\delta W[J,K]}{\delta J_a(x)}\,,\label{eq:mean_field}\\
    \hbar\,\Delta^{ab}(x,y)\,&\equiv\,2\,\frac{\delta W[J,K]}{\delta K_{ab}(x,y)}\,-\,\phi^a(x)\,\phi^b(y)\,,\label{eq:2point}
\end{align}
where we identify $\phi$ and $\Delta$ with the connected one- and two-point correlation functions. 
To be precise, we stress here that $\phi$, also referred to as the \textit{mean field}, is defined as the expectation value of the full quantum field $\Phi$ in the presence of the sources $J$ and $K$. $\Delta$ is the full propagator, i.e., the connected two-point function computed from the full path integral:\ $\Delta^{ab} \,=\, \langle \mathrm{T}[\Phi^a \,\Phi^b]\rangle_{J,K} -\langle \Phi^a \rangle_{J,K}\, \langle\Phi^b\rangle_{J,K} $. 

\begin{figure}
\centering
    \begin{subfigure}{0.28\textwidth}
        \centering
        \includegraphics[width=\linewidth]{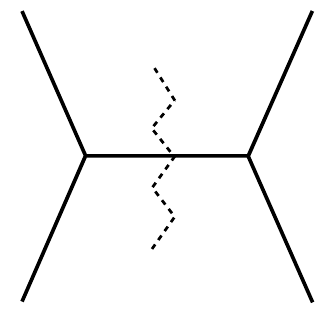}
        \caption{}
        \label{fig:reducible-a}
    \end{subfigure}
    \hspace{2cm}
    \begin{subfigure}{0.28\textwidth}
        \centering
        \includegraphics[width=\linewidth]{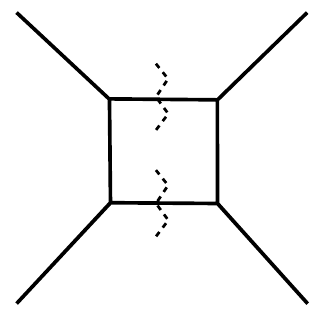}
        \caption{}
        \label{fig:reducible-b}
    \end{subfigure}    
\caption{Examples of reducible and irreducible diagrams. Figure \ref{fig:reducible-a} displays a one-particle-reducible diagram, with the corresponding internal line cut indicated explicitly. Figure \ref{fig:reducible-b} shows the so-called box diagram, which provides an example of a one-particle-irreducible diagram that is nevertheless two-particle reducible.} \label{fig:reducible}
\end{figure}

Given that the tangents to the Schwinger functional are the correlation functions, performing the Legendre transform of $W[J,K]$ allows us to exchange the description in terms of sources $J,K$ for one in terms of correlators $\phi, \Delta$. This is precisely what we want to do. Thus, we define the quantum effective action
\begin{equation}
    \label{eq:2PIstart}
    \Gamma[\phi,\Delta]\,=\,{\rm max}_{J,K}\left[- W[J,K]\,+\int \mathrm{d}^4 x\, J_a(x)\,\phi^a(x)\,+\,\frac{1}{2}\int \mathrm{d}^4 x\,\mathrm{d}^4 y\,K_{ab}(x,y)\,\big(\phi^a(x)\,\phi^b(y)\,+\,\hbar\, \Delta^{ab}(x,y)\big)\right]\,.
\end{equation}
In fact, this is the so-called \emph{two-particle-irreducible} or \emph{2PI effective action}, due to Cornwall, Jackiw and Tomboulis~\cite{Cornwall:1974vz}.
Its interpretation is most clear by making use of the diagrammatic language. An $n$-particle-irreducible ($n$PI) diagram is defined as a connected diagram that remains connected upon cutting any set of up to $n$ internal propagator lines. \emph{Irreducible} diagrams therefore constitute the fundamental building blocks of quantum dynamics, encoding genuinely non-factorizable processes. \emph{Reducible} diagrams can be constructed by iteratively ``gluing'' together irreducible parts. An illustrative comparison between reducible and irreducible diagrams is shown in Figure \ref{fig:reducible}.

Before going into the details of all things effective action, a word of caution may be in order. Here, we choose to take a different approach to tackling this topic compared to the standard QFT textbook. The textbook prescription is to study only a particular limit of the $2$PI effective action, namely the $K\rightarrow 0$ limit, which gives rise to the $1$PI effective action. Instead, here we choose to start from the often, and unlawfully, disregarded $2$PI functional, which is in fact an excellent candidate to be studied in geometric terms. But worry not:\ we will still discuss the $1$PI limit and its geometric structure in \ref{sec:1pi}.

Before getting back to where we left off, and having got rather tired of writing so many integrals, we will now use a DeWitt-like notation in which repeated indices and coordinate arguments are summed and integrated over, respectively. For example, 
\begin{equation}\label{eq:dewitt_convention}
    J_a\,\phi^a\,\Longrightarrow\, \sum_a\int{\rm d}^4x\,J_a(x)\,\phi^a(x)\,.
\end{equation}

After the extremization, we have particular sources $J$ and $K$, which are those that maximize the functional in square brackets in~\eqref{eq:2PIstart}. Thus, the extremization defines $J\,=\,J[\phi,\Delta]$ and $K\,=\,K[\phi,\Delta]$ to be functionals of $\phi$ and $\Delta$, in much the same way as the Legendre transform between Lagrangian and Hamiltonian mechanics makes the momentum $p$ a function of the time derivative of the generalized coordinate $q$, i.e., $p\,=\,p(\dot{q})$. With this understood, and with a slight abuse of notation for the sources, we can drop the extremization and write the quantum effective action
\begin{equation}\label{eq:2pi_ea}
    \Gamma[\phi,\Delta]\,=\,-\, W[J,K]\,+\,J_a\,\phi^a\,+\,\frac{1}{2}K_{ab}\,\big(\phi^a\,\phi^b\,+\,\hbar \,\Delta^{ab}\big)\,.
\end{equation}
We now compute the functional derivatives of the effective action with respect to the connected one- and two-point functions, obtaining
\begin{align}
    \frac{\delta \Gamma[\phi,\Delta]}{\delta \phi^a}&=J_a\,+\,K_{ab}\,\phi^b\;,\\
    \frac{\delta \Gamma[\phi,\Delta]}{\delta \Delta^{ab}}&=\frac{1}{2}K_{ab}\;.
\end{align}
Noticing the presence of the sources on the right-hand side, we infer that these relations are nothing other than the equations of motion for the one- and two-point functions. They are, however, not the classical equations of motion; they are the \textit{quantum-corrected} equations of motion, wherein quantum fluctuations have been integrated out to give effective classical-number differential equations for the correlation functions. And this brings us to an important facet of the question, ``how do we build theories?'':\ \emph{integrating stuff out}.

We are probably all familiar with the maxim, ``Sociology is Biology, Biology is Chemistry, Chemistry is Physics, Physics is Maths, and no one knows what Maths is.'' As crass as this statement is, it is a reflection of what we might call \emph{constructionism}, which is often and incorrectly assumed to be the corollary of reductionism. To be similarly crass:\ knowing all the physics that there is to know won't help you one bit to explain why certain species of birds are so partial to a mate who can moonwalk. The reason is that the degrees of freedom are all wrong. Trying to understand the mating rituals of the world's fauna in terms of the dynamics of elementary particles is a fool's errand. We might like to think that the rules of social engagement can be derived from fundamental physics, but \emph{complexity} has taught us otherwise.

But we can consider far less extreme hierarchies. For example, we know that the most useful degrees of freedom for describing QCD at high energies are quarks and gluons and, at low energies, they are hadrons. What we learn from this is that the most effective way of describing the physics can depend on the characteristic energy or length scale of the system of interest. We move from shorter length scales to larger ones by averaging over effects on smaller scales. We call this \emph{coarse graining}. In quantum field theory, this process amounts to integrating out quantum fluctuations on particular scales and this is, as mentioned earlier, at the heart of renormalization.

Looking again at the 2PI effective action, we notice that the source $K(x,y)$ is accompanied by two factors of the field in the path integral. Thus, by an appropriate choice of $K$, we can make the field more or less massive. Then, since the source $K$ is coordinate dependent, its Fourier transform $\tilde{K}(p)$ can be used to make certain Fourier modes and therefore certain quantum fluctuations of the field more or less massive. This allows us to re-weight contributions from particular fluctuations, e.g., freezing those with momentum $p$ less than some scale $k$, such that only those with $p>k$ are integrated out when evaluating the partition function. In this way, the Fourier transform of the two-point source can be used to implement a smooth cutoff, which serves as a \emph{regulator}. We can then ask how the Schwinger functional changes as a function of the scale $k$ -- this defines the \emph{renormalization group} or \emph{RG evolution} (see, e.g., Ref.~\cite{Tong}) --- and we will return to this shortly.

To start with, we are going to integrate out \emph{all} the quantum fluctuations to get our bearings on the form of the 2PI effective action.  This will be enough for us to see how the resulting effective dynamics can differ from those encoded in the classical action.  Let's start with the case $N\,=\,1$ and impose some additional symmetry on the classical action to constrain the number of operators. To this end, we impose $\mathbb{Z}_2$ symmetry, so that the action is symmetric under $\Phi\to\,-\,\Phi$. To keep things simpler still, we will do everything in Euclidean space,  neglect derivative interactions, and stop at operators quartic in the field $\Phi$. Thus, we take the partition function
\begin{equation}\label{eq:partition_func}
Z[J,K]\,=\,\mathcal{N}\int\left[\mathrm{d}\Phi\right]\exp\left[\,-\,\frac{1}{\hbar}\left(S[\Phi]\,-\,J(x)\,\Phi(x)\,-\,\frac{1}{2}K(x,y)\,\Phi(x)\,\Phi(y)\right)\right]\,,
\end{equation}
with classical action
\begin{equation}
    S[\Phi]\,=\,\int{\rm d}^4x\,\left(\frac{1}{2}Z_{\Phi}\,\big(\partial\Phi(x)\big)^2\,+\,\frac{1}{2}\,m^2\,\Phi^2(x)\,+\,\frac{\lambda}{4!}\,\Phi^4(x)\right)\,,
\end{equation}
and we fix $m^2\,>\,0$ and $\lambda\,>\,0$.
Integration over repeated spacetime arguments is understood in \eqref{eq:partition_func} and in all subsequent expressions, according to the DeWitt convention introduced in \eqref{eq:dewitt_convention}.

So far, we have allowed for a general wavefunction factor $Z_{\Phi}$. However, by making the following redefinitions of the fields and sources, we can eliminate $Z_{\Phi}$ from the action:
\begin{align}
    \Phi(x)\,\longrightarrow\, Z_{\Phi}^{\,-1/2}\,\Phi(x)\,,\qquad
    J(x)\,\longrightarrow\, Z_{\Phi}^{1/2}\,J(x)\,,\qquad
    K(x,y)\,\longrightarrow\, Z_{\Phi}\,K(x,y)\,.
\end{align}
This is an important observation about our charting of the theory space in terms of parameters (or operators):\ some parameters (and indeed operators) can be removed by field redefinitions. We call the corresponding operators \emph{inessential} (see~\cite{Baldazzi:2021ydj}), and they are a redundant part of our mathematical description that does not impact physical observables. Making these redefinitions, the classical action becomes
\begin{equation}
    S[\Phi]\,=\,\int{\rm d}^4x\left(\frac{1}{2}\big(\partial\Phi(x)\big)^2\,+\,\frac{1}{2}\,m^2\,\Phi^2(x)\,+\,\frac{\lambda}{4!}\,\Phi^4(x)\right)\,,
\end{equation}
and thankfully we no longer have to worry about getting our $Z$'s confused.

The functional integral in the partition function will be dominated by configurations that extremize the exponent. Our aim is to find these configurations and perform a so-called \emph{saddle-point approximation}. We define the saddle point $\varphi$ as the field that satisfies
\begin{equation}
    \label{eq:stationarity}
    \frac{\delta S[\Phi]}{\delta \Phi(x)}\bigg|_{\Phi\,=\,\varphi}\,=\,J(x)\,+\,K(x,y)\,\varphi(y)\,.
\end{equation}
Hence, $\varphi$ is defined by the stationarity condition of the full functional integral $Z$ at fixed sources $J,K$. This is formally different from the mean field $\phi$ defined in \eqref{eq:mean_field} and in general $\phi \,\neq\, \varphi$. We proceed by writing $\Phi(x)\,=\,\varphi(x)\,+\,\hbar^{1/2}\,\delta\Phi(x)$, where the field $\delta\Phi(x)$ encodes the quantum fluctuations and $\hbar^{1/2}$ is for bookkeeping purposes. The classical action is then expanded as
\begin{align}
S[\Phi]\,&=\,S[\varphi]\,+\,\hbar^{1/2}\,\frac{\delta S[\Phi]}{\delta \Phi(x)}\bigg|_{\Phi\,=\,\varphi}\,\delta\Phi(x)\,+\,\frac{\hbar}{2!}\,\delta \Phi(x)\,\frac{\delta^2 S[\Phi]}{\delta \Phi(x)\,\delta\Phi(y)}\bigg|_{\Phi\,=\,\varphi}\,\delta \Phi(y)\nonumber\\&\phantom{=}\,+\,\frac{\hbar^{3/2}}{3!}\,S^{(3)}(\varphi;x)\,\delta\Phi^3(x)\,+\,\frac{\hbar^{2}}{4!}S^{(4)}(\varphi;x)\,\delta\Phi^4(x)\,,
\end{align}
where
\begin{align}
    S^{(3)}(\varphi;x)\,&=\,\int{\rm d}^4y\,{\rm d}^4z\,\frac{\delta^3 S[\Phi]}{\delta \Phi(x)\,\delta\Phi(y)\,\delta\Phi(z)}\bigg|_{\Phi\,=\,\varphi}\,=\,\lambda\,\varphi(x)\,,\\
    S^{(4)}(\varphi;x)\,&=\,\int{\rm d}^4y\,{\rm d}^4z\,{\rm d}^4w\,\frac{\delta^4 S[\Phi]}{\delta \Phi(x)\,\delta\Phi(y)\,\delta\Phi(z)\,\delta\Phi(w)}\bigg|_{\Phi\,=\,\varphi}\,=\,\lambda\,.
\end{align}
The partition function can be written
\begin{align}
    Z[J,K]\,&=\,\mathcal{N}\,\exp\left[\,-\,\frac{1}{\hbar}\,\left(S[\varphi]\,-\,J(x)\varphi(x)\,-\,\frac{1}{2}\,K(x,y)\,\varphi(x)\,\varphi(y)\right)\right]\nonumber\\&\phantom{=}\times\int\left[\mathrm{d}\,\delta \Phi\right]\,\exp\left[-\frac{1}{2}\,\delta\Phi(x)\,\mathcal{G}^{-1}(x,y)\,\delta\Phi(y)\,-\,\frac{\hbar^{1/2}}{3!}\,S^{(3)}(\varphi;x)\,\delta\Phi^3(x)\,-\,\frac{\hbar}{4!}\,S^{(4)}(\varphi;x)\,\delta\Phi^4(x)\right]\,,
\end{align}
where we have introduced the inverse two-point function
\begin{equation}\label{eq:dyson}
    \mathcal{G}^{-1}(x,y)\,=\,G^{-1}(\varphi;x,y)\,-\,K(x,y)\,=\,\frac{\delta^2 S[\Phi]}{\delta \Phi(x)\,\delta \Phi(y)}\bigg|_{\Phi\,=\,\varphi}\,-\,K(x,y)\,.
\end{equation}
A little pedantry is needed here:\ we must keep in mind that the sources $J$ and $K$ are still functionals of the one- and two-point functions $\phi$ and $\Delta$.

Assuming that $S^{(3)}$ and $S^{(4)}$ are controlled by small parameters, we can compute the functional integral over the fluctuations perturbatively. The leading term is
\begin{equation}
    \int\left[\mathrm{d}\,\delta \Phi\right]\,\exp\left[-\,\frac{1}{2}\,\delta\Phi(x)\,\mathcal{G}^{-1}(x,y)\,\delta\Phi(y)\right]\,\propto \,\mathrm{det}^{-1/2}\,\mathcal{G}^{-1}\,,
\end{equation}
wherein we have omitted constant factors. In Euclidean space, the Schwinger functional is
\begin{equation}
    W[J,K]\,=\,\hbar\,\ln Z[J,K]
\end{equation}
and the quantum effective action up to order $\hbar^2$ can be written as
\begin{align}
    \Gamma[\phi,\Delta]\,&=\,\Gamma_0[\varphi]\,+\,\hbar\,\Gamma_1[\varphi,\mathcal{G}]\,+\,\hbar^2\,\Gamma_2[\varphi,\mathcal{G}]\,+\,\hbar^2\,\Gamma_{\rm 1PR}[\varphi,\mathcal{G}]\nonumber\\&\phantom{=}\,+\,\big(J(x)\,+\,K(x,y)\,\phi(y)\big)\,\big(\phi(x)\,-\,\varphi(x)\big)\nonumber\\&\phantom{=}\,-\,\frac{1}{2}\,K(x,y)\,\big[\big(\phi(x)\,-\,\varphi(x)\big)\,\big(\phi(y)\,-\,\varphi(y)\big)\,-\,\hbar\big(\Delta(x,y)\,-\,\mathcal{G}(x,y)\big)\big]\,,
\end{align}
or, more neatly, as
\begin{align}
    \Gamma[\phi,\Delta]\,&=\,\Gamma_0[\varphi]\,+\,\hbar\,\Gamma_1[\varphi,\mathcal{G}]\,+\,\hbar^2\,\Gamma_2[\varphi,\mathcal{G}]\,+\,\hbar^2\,\Gamma_{\rm 1PR}[\varphi,\mathcal{G}]\nonumber\\&\phantom{=}\,+\,\frac{\delta \Gamma[\phi,\Delta]}{\delta \phi(x)}\,\big(\phi(x)\,-\,\varphi(x)\big)\nonumber\\&\phantom{=}\,-\,\frac{1}{\hbar}\,\frac{\delta \Gamma[\phi,\Delta]}{\delta \Delta(x,y)}\,\big[\big(\phi(x)\,-\,\varphi(x)\big)\,\big(\phi(y)\,-\,\varphi(y)\big)\,-\,\hbar\,\big(\Delta(x,y)\,-\,\mathcal{G}(x,y)\big)\big]\,,
\end{align}
where
\begin{align}\Gamma_0[\varphi]\,&=\,S[\varphi]\,, \\ \Gamma_1[\varphi,\mathcal{G}]\,&=\,\frac{1}{2}\mathrm{tr}\Big[\ln\big(\mathcal{G}^{\,-1} \ast G(0)\big)\,+\,G^{\,-1}(\varphi)\ast \mathcal{G}\,-\,1\Big]\,,\\
\Gamma_2[\varphi,\mathcal{G}]\,&=\, -\,\raisebox{0.1\height}{\includegraphics[valign=c, scale=0.5]{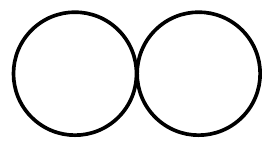}}\,-\,\raisebox{0.06\height}{\includegraphics[valign=c, scale=0.55]{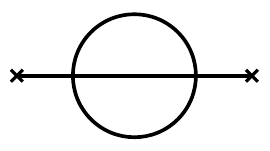}}\,,
\\ 
\Gamma_{\mathrm{1PR}}[\varphi,\mathcal{G}]\,&=\,-\, \raisebox{0.1\height}{\includegraphics[valign=c, scale=0.55]{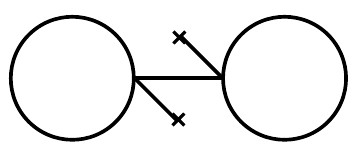}}\,.
\end{align}
The $\ast$ indicates that the two-point functions are integrated over the volume, and we have introduced a normalization $G(0)$ to make the logarithm dimensionless. $\Gamma_2$ and $\Gamma_{\rm 1PR}$ contain 2PI and one-particle reducible (1PR) diagrams, respectively. The cross symbol attached to the external legs denotes field insertions. 

If we assume that the path integral is dominated by a single saddle-point configuration --- this is not actually true for general $J(x)$ and $K(x,y)$ --- we can eliminate $\phi$ and $\Delta$ for $\varphi$ and $\mathcal{G}$ (see~\cite{Garbrecht:2015cla, Millington:2019nkw}).  This is not just a futile exercise made to do more pages of algebra. Instead, in this case, $\varphi$ and $\mathcal{G}$ will turn out to be the physical configurations (cf. \eqref{eq:dyson}).

To this end, we expand
\begin{align}
    \Gamma[\phi,\Delta]\,&=\,\Gamma[\varphi,\mathcal{G}]\,+\frac{\delta \Gamma[\phi,\Delta]}{\delta \phi(x)}\bigg|_{\substack{\phi\,=\,\varphi\\ \Delta\,=\,\mathcal{G}}}\,\big(\phi(x)-\varphi(x)\big)\,+\,\frac{1}{2}\big(\phi(x)\,-\,\varphi(x)\big)\,\frac{\delta^2\Gamma[\phi,\Delta]}{\delta \phi(x)\delta \phi(y)}\bigg|_{\substack{\phi\,=\,\varphi\\ \Delta\,=\,\mathcal{G}}}\,\big(\phi(y)\,-\,\varphi(y)\big)\nonumber\\&\phantom{=}\,+\,\frac{\delta \Gamma[\phi,\Delta]}{\delta \Delta(x,y)}\bigg|_{\substack{\phi\,=\,\varphi\\ \Delta\,=\,\mathcal{G}}}\,(\Delta(x,y)\,-\,\mathcal{G}(x,y))\,+\,\mathcal{O}(\hbar^3)\,.
\end{align}
Similarly, we expand
\begin{align}
    \frac{\delta \Gamma[\phi,\Delta]}{\delta \phi(x)}\,&=\,\frac{\delta \Gamma[\phi,\Delta]}{\delta \phi(x)}\bigg|_{\substack{\phi\,=\,\varphi\\ \Delta\,=\,\mathcal{G}}}\,+\,\frac{\delta^2 \Gamma[\phi,\Delta]}{\delta \phi(x)\delta \phi(y)}\bigg|_{\substack{\phi\,=\,\varphi\\ \Delta\,=\,\mathcal{G}}}\,\big(\phi(y)\,-\,\varphi(y)\big)\,+\,\mathcal{O}(\hbar^2)\,,\\
    \frac{\delta \Gamma[\phi,\Delta]}{\delta \Delta(x,y)}\,&=\,\frac{\delta \Gamma[\phi,\Delta]}{\delta \Delta(x,y)}\bigg|_{\substack{\phi\,=\,\varphi\\ \Delta\,=\,\mathcal{G}}}\,+\,\mathcal{O}(\hbar^2)\,.
\end{align}
Putting everything together, we are left with (at order $\hbar^2$)
\begin{align}
    \Gamma[\phi,\Delta]\,&=\Gamma_0[\varphi]\,+\,\hbar\Gamma_1[\varphi,\mathcal{G}]\,+\,\hbar^2\Gamma_2[\varphi,\mathcal{G}]\,+\,\hbar^2\Gamma_{\rm 1PR}[\varphi,\mathcal{G}]\nonumber\\&\phantom{=}\,+\,\frac{1}{2}\big(\phi(x)\,-\,\varphi(x)\big)\left[\frac{\delta^2 \Gamma[\phi,\Delta]}{\delta \phi(x)\delta \phi(y)}\bigg|_{\substack{\phi\,=\,\varphi\\ \Delta\,=\,\mathcal{G}}}\,-\,\frac{2}{\hbar}\frac{\delta\Gamma[\phi,\Delta]}{\delta \Delta(x,y)}\right]\big(\phi(y)\,-\,\varphi(y)\big)\,.
\end{align}
The terms in the second line cancel, eliminating the 1PR diagrams, and, after showing an admirable level of patience, we arrive at the final expression
\begin{equation}
    \Gamma[\varphi,\mathcal{G}]\,=\,\Gamma_0[\varphi]\,+\,\hbar\Gamma_1[\varphi,\mathcal{G}]\,+\,\hbar^2\Gamma_2[\varphi,\mathcal{G}]\,.
\end{equation}

Notice that the equation of motion for $\varphi$ can be derived in two ways: via the partial functional derivative of $\Gamma[\varphi,\mathcal{G}]$ with respect to $\varphi$, or from the stationarity condition~\eqref{eq:stationarity}. For these to be consistent, it is clear that we require
\begin{equation}
\frac{\Gamma[\phi,\Delta]}{\delta \phi(x)}\bigg|_{\substack{\phi\,=\,\varphi\\ \Delta\,=\,\mathcal{G}}}\,=\,\Big[J(x)\,+\,K(x,y)\phi(y)\Big]_{\substack{\phi\,=\,\varphi\\ \Delta\,=\,\mathcal{G}}}\,=\,0\,,
\end{equation}
and we can show it to be the case if we recall that $J$ and $K$ are functionals of $\phi$ and $\Delta$. This approach is particularly useful when the extremum of the quantum effective action and the extremum of the classical action differ by more than some small amount~\cite{Garbrecht:2015cla}. Thus, $\varphi$ satisfies the following quantum-corrected equation of motion at order $\hbar$:
\begin{equation}
\begin{aligned}
    \frac{\delta\Gamma[\varphi,\mathcal{G}]}{\delta \varphi(x)}\,&=\,\frac{\delta S[\varphi]}{\delta \varphi(x)}\,+\,\frac{\hbar}{2}\frac{\delta G^{-1}(\varphi)}{\delta \varphi(x)}\,\mathcal{G}=\,\left(-\,\partial^2\varphi(x)\,+\,m^2\varphi(x)\right)\,+\,\frac{\lambda}{3!}\varphi^3(x)\,+\,\frac{\hbar\lambda}{2}\varphi(x)\,\mathcal{G}(x,x)\,,\\
    &=\; \raisebox{0.04\height}{\includegraphics[valign=c, scale=0.55]{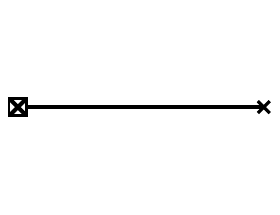}}\;+\;\raisebox{0.03\height}{\includegraphics[valign=c,   scale=0.55]{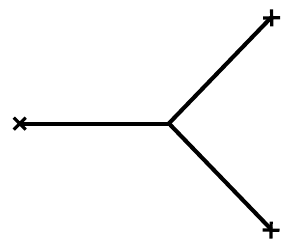}}\;+\;\raisebox{0.05\height}{\includegraphics[valign=c, scale=0.6]{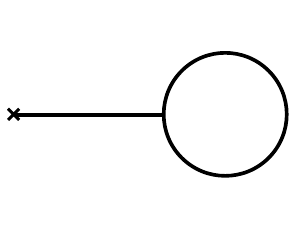}}\;,
\end{aligned}
\end{equation}
wherein we recognize the third term on the right-hand side as the one-loop tadpole proportional to $\mathcal{G}(x,x)$. (The first diagram corresponds to the terms in parentheses, and the second term to the term cubic in the field $\varphi$.)

The 2PI effective action is a useful hammer, and it gives us a powerful organizing principle for the resummation of \textit{loop} corrections. We can also use it to understand how theories change with energy scale via the exact renormalization group, as suggested earlier. %A case of particular interest is the $K\to 0$ limit which allows us to obtain the 1PI effective action from its 2PI counterpart. 
In the next section, we will uncover an insightful geometric interpretation of the effective action beyond that provided by the Legendre transforms discussed earlier. 

\section{The geometry of theory space}

In the last section, we made a sightseeing tour of the 2PI effective action. Let us now take a step back and study the 1PI effective action in more detail.

\subsection{Geometric 1PI}\label{sec:1pi}
As mentioned in the previous section, the 1PI effective action can be conveniently obtained from the 2PI expression by taking the limit of vanishing two-point source, $K\to 0$, in \eqref{eq:2pi_ea}:
\begin{equation}\label{eq:1pi_ea}
    \Gamma_{\mathrm{1PI}}[\phi]=\,-\, W[J]\,+\,J_a\phi^a.
\end{equation}
The one- and two- point functions are then defined according to \eqref{eq:mean_field} and \eqref{eq:2pt_1pi}, noting that the Schwinger functional now depends only on $J$. 

Differentiating \eqref{eq:1pi_ea} with respect to $J_a$ and substituting the result into \eqref{eq:1p1_func} yields an implicit equation for $ \Gamma_{\mathrm{1PI}}[\phi]$:
\begin{equation}\label{eq:1p1_implicit}
  \exp \left(\frac{i}{\hbar} \Gamma_{\mathrm{1PI}}[\phi]\right)\,=\int \left[\mathrm{d} \Phi\right] \exp \left(\frac{i}{\hbar} S\left[\Phi\right]\,+\,\frac{i}{\hbar} \frac{\delta\Gamma_{\mathrm{1PI}}[\phi]}{\delta \phi^a} \left(\phi^a\,-\, \Phi^a\right)\right)\,. 
\end{equation}
Performing the semiclassical expansion $\Phi\,=\,\varphi\,+\,\hbar^{1/2}\delta\Phi$, allows us to write \eqref{eq:1p1_implicit} perturbatively as 
\begin{equation}\label{eq:loop_sgqea}
    \Gamma[\varphi] \,=\, \sum_{l\,=\,0} \hbar^l \,\Gamma_{l}[\varphi]\,,
\end{equation}
where $\Gamma_{0}[\varphi]\,=\,S[\varphi]$ as per usual and, from now on, we set $\hbar\,=\,1$. With this in mind,  \eqref{eq:1p1_implicit} can be solved order by order in perturbation theory. Before continuing, a (not so) brief digression is in order. 

In our journey toward building a quantum field theory, we noted that we should have full freedom in how we chart the theory space. The choice of fundamental variables is an arbitrary one, and we would naturally expect observables to be independent of this choice. However, this turns out to be a more complicated matter than one might naively expect. We mentioned earlier in this lecture that the operators of a theory may be viewed as coordinates, and that some of the parameters associated with these operators can turn out to be redundant. We now adopt this perspective as our starting point by treating the fields $\Phi^a$ as chart variables of an infinite-dimensional  \textit{configuration-space manifold}.
 
We may now wish to express our quantum field theory in terms of another set of fields, say $\widetilde{\Phi}$. We refer to this exercise as a \textit{field redefinition} or \textit{field reparameterization}. This operation can be written as 
\begin{equation}
   \label{eq:PhiDiff}
    \Phi^{a}\ \rightarrow\ \widetilde{\Phi}^{a}\,=\, \widetilde{\Phi}^{a}(\Phi)\,.
\end{equation}
In this lecture, we consider a small subset of all the possible field transformations \eqref{eq:PhiDiff}, namely those field redefinitions that do not introduce extra spacetime derivatives of the fields like $\partial \Phi$. (These are also termed \textit{ultralocal} in the literature \cite{Gattus:2024ird,Gattus:2023gep}.)  Simply speaking, in QFT, a function or operator is said to be local if its value at one spacetime point depends solely on field configurations evaluated at the same point, or at most infinitesimally close to that point. Working with this restricted subset of allowed field redefinitions means that we can avoid issues with causality in propagating degrees of freedom while still covering a wide variety of physical theories.

Now that we have a manifold spanned by fields, we want to define a metric on it. Just as spacetime can be interpreted as a manifold with a spacetime metric, we want a means to ``feel'' distances between different points of the configuration-space manifold. We can imagine defining a set of ``local rulers'' at each point. Even with local information about the space near each point, it would be better if we could somehow view the manifold from a distance to understand its overall structure. In the language of differential geometry, this information is encoded in the \textit{global metric} $g_{ab}(\Phi)$.

To find the analytic form of the metric, we consider our simple scalar field theory with $N$ fields, described by the general action functional \eqref{eq:action}, written here using the DeWitt notation:
\begin{equation}\label{eq:action_geometry}
S[\Phi]\,=\,\eta^{\mu\nu}\,Z_{ab}(\Phi)\,\partial_{\mu}\Phi^a\, \partial_{\nu}\Phi^b\,-\,V(\Phi)\,.
\end{equation}
Let us analyze the elements appearing in the action above through our new geometrical lens. Interpreting $\Phi^a$ as coordinates, $\partial_{\mu}\Phi^a$ becomes a configuration-space vector (in addition to a spacetime covector), $Z_{ab}(\Phi)$ is a rank-2 tensor, and $V(\Phi)$ transforms as a scalar under \eqref{eq:PhiDiff}, meaning that $\widetilde{V}(\widetilde{\Phi})= V(\Phi(\widetilde{\Phi}))$. We would like the configuration-space manifold to be pseudo-Riemannian so that it inherits some convenient geometric properties. The global metric should satisfy three key properties: (i) it is a rank-2 tensor under field redefinitions; (ii) it can be derived from the action functional; and (iii) it is flat for canonically normalized theories. Property (i) is a standard requirement for any manifold metric, whereas properties (ii) and (iii) are specific to configuration spaces. Property (ii) ensures that the geometry of the configuration space is fully determined by the underlying quantum field theory, without requiring any additional structures.  Finally, the last property ensures that, for a free field theory, the metric reduces to the identity matrix, $g_{ab}(\Phi)\,=\,\delta_{ab}$.

As it turns out, we are in luck: the scalar-field action in \eqref{eq:action_geometry} already contains a quantity satisfying properties (i)--(iii): the kinetic matrix $Z$. Hence, we identify $g_{ab}(\Phi)~=~Z_{ab}(\Phi)$ and note that, for $S[\Phi]$ as above, the metric can be uniquely extracted via:
\begin{equation}
   \label{eq:metric}
g_{ab}\ =\ \frac{\eta_{\mu \nu}}{4} \frac{\partial^2 S}{\partial(\partial_{\mu} \Phi^{a})\,\partial(\partial_{\nu} \Phi^{b})}\,.
\end{equation}
Obtaining the metric from the action as above ensures that the configuration space ``knows'' about the specific theory we are considering.  In our case, this information is provided by the quadratic form multiplying spacetime derivatives of the fields. The potential $V(\Phi)$ is a scalar function under \eqref{eq:PhiDiff}, and as such it contains no information about the notion of distances, angles, or kinetic inner products. In fact, it is $g_{ab}$, as obtained through \eqref{eq:metric}, which tells us how to write down a covariant line element $\mathrm{d} s^2= g_{ab}\, \mathrm{d}\Phi^a\, \mathrm{d}\Phi^b$, in analogy with the invariant interval $\mathrm{d} s^2= g_{\mu \nu}\, \mathrm{d}x^\mu\, \mathrm{d}x^\nu$ defining the infinitesimal spacetime distance between two events in GR. As it stands, the potential $V(\Phi)$ can therefore affect the dynamics of the fields on configuration space, but not the metric geometry itself.

Now, one could argue that a quadratic object could also be defined from the second variation of the potential, leading to a contribution of the type $\partial_a\partial_b V \, \mathrm{d}\Phi^a\, \mathrm{d}\Phi^a$. However, $\partial_a\partial_b V$ does not transform as a proper tensor under \eqref{eq:PhiDiff}. We know from GR that a proper tensor can only be constructed by means of covariant differentiation, i.e. $\nabla_a\nabla_b V(\Phi)$.\footnote{\begingroup
\setlength{\abovedisplayskip}{6pt}
\setlength{\belowdisplayskip}{6pt}
\setlength{\abovedisplayshortskip}{6pt}
\setlength{\belowdisplayshortskip}{6pt}
Recall that the covariant derivative $\nabla$ of a contravariant vector field $X^a$ is defined as 
\begin{equation*}
    \nabla_b X^a \,=\, \partial_b X^a \,+\, \Gamma^a_{\:bc}\,X^c,
\end{equation*}
while the Riemann tensor is defined to be
\begin{equation*}
    R^{a}_{\:\:bcd}\,= \,-\,\partial_d\Gamma_{\:\:bc}^a\,+\, \partial_c\Gamma_{\:\:bd}^a\,+\,\Gamma_{\:\:mc}^a \,\Gamma_{\:\:bd}^m \,-\,\Gamma^a_{\:\:md}\, \Gamma^m_{\:\:bc}\,.
\end{equation*}\endgroup} 

However, building this object already requires knowing the specific form of the metric connection, which in turn requires knowing the metric. That is not to say that the quadratic form derived from the potential is of no use; quite the opposite, in fact (spoiler alert: we will learn more about this in \ref{sec:geometric_2pi}).

If one accepts in good conscience that the covariant metric is the one given in \eqref{eq:metric}, it is also imperative to specify that the profile of the so-defined configuration space is dictated by the \textit{quadratic derivative interactions}, while the non-derivative interactions are relegated to $V(\Phi)$.  Notice then that the $N\,=\,1$ case is therefore akin to free field theory with respect to the geometry. Since there is only one field species, the metric $g_{ab}$ reduces to $f(\phi)\,\delta_{ab}$. Thus, while we could still have a non-canonical kinetic term, this can be brought to canonically normalized form by a suitable field redefinition. Since the classical action is invariant under field reparameterizations, i.e. $S[\Phi]\,=\,\widetilde{S}[\widetilde{\Phi}]$, we are always allowed to perform such a change of variables. When considering multiple fields instead, we can render the kinetic term canonically normalized only if the configuration space is trivial, i.e., flat. Now let's go back to $N$ fields.

Once the metric is defined, we can borrow familiar tools from differential geometry, such as the covariant (functional) derivative $\nabla$ and the curvature tensor $R^{a}_{\:\:bcd}$, to compute \textit{covariant} interaction vertices.  We follow the notation first introduced by Ecker and Honerkamp \cite{Ecker:1972tii}, which yields conveniently concise expressions. Taking the covariant functional derivative of $\partial_\mu \Phi^a(x)$ with respect to another field $\Phi^b(y)$, we define the operator 
\begin{equation}
  \label{eq:Dmuab}
    \nabla_{\Phi^b}\,\partial_\mu \Phi^a\,=\,\partial_{\mu}\,\delta^a_{\:\:b} \,+\,\Gamma^{a}_{\:\:bc}\:\partial_\mu\Phi^c \,\equiv(D_\mu)^a_{\:\:b}\,,
\end{equation}
where $\Gamma^{a}_{\:\:bc}$ are the components of the configuration-space connection. The convenience of this choice becomes evident once we take a further covariant derivative of $D_\mu$:
\begin{equation}
   \label{eq:DmuRabc}
     \nabla_{\Phi^c}\,(D_\mu)_{ab}\,=\,R_{abcd}\,\partial_\mu\Phi^d \,.
\end{equation}
We can now proceed to compute fully \textit{off-shell} (we will return to this shortly) covariant expressions for the scalar interaction vertices (see \cite{Gattus:2024ird} and references therein). We can derive the classical equation of motion for the fields $\Phi^a$ as
\begin{equation}
   \label{eq:Sa}
    \frac{\delta S}{\delta \Phi^a}\ =\ \eta^{\mu\nu}\,(\partial_\mu\Phi^m)\, (D_\nu)_{ma}\,-\, \frac{\delta V}{\delta \Phi^a}\,.
\end{equation}
The covariant inverse propagator is found to be:
\begin{equation}
   \label{eq:Sab}
\nabla_b \nabla_a S\,=\,(D_\mu)^m_{\:\:a}\:(D^\mu)_{mb}\,+\,\,\partial_\mu\Phi^m\,R_{mabp}\:\partial^\mu\Phi^p \,-\,\nabla_b \nabla_aU\,.
\end{equation}
Similarly, higher-point interaction vertices can be obtained by applying additional covariant derivatives to $S[\Phi]$; we will not do that here. Instead, we return to the question of the 1PI effective action in  \eqref{eq:1p1_implicit}. 

Vilkovisky \cite{Vilkoviskii:1984un} first pointed out that the standard definition of $\Gamma_{\mathrm{1PI}}[\phi]$ is not strictly invariant under field reparameterizations because of the $\left(\phi^a\,-\, \Phi^a\right)$ term. This is just the difference between two coordinates and does not transform as a proper configuration-space vector under \eqref{eq:PhiDiff}. But no need to panic just yet. The non-covariance of the standard effective action does not imply that our physical observables, obtained in the \textit{on-shell} limit, are not reliable. In fact, we would only get distinct results in the \textit{off-shell} regime. Looking back at \eqref{eq:1p1_implicit}, one can show that the difference between the effective action $\Gamma$ and a new functional $\widetilde{\Gamma}$, obtained via an ultralocal field redefinition $\Phi \rightarrow \widetilde{\Phi}$, is proportional to the first variation of the classical action, ${\delta S}/{\delta \Phi^a}$. When external sources vanish and the classical equations of motion are satisfied, on-shell quantities, such as the $S$-matrix, are unaffected by the particular field parameterization.

Vilkovisky suggested replacing the non-covariant term $\left(\phi^a\,-\, \Phi^a\right)$ in \eqref{eq:1p1_implicit} with a new configuration-space vector $\sigma^{a}\left[\phi,\!\Phi\right]$, defined as the geodesic tangent vector connecting $\phi$ to $\Phi$, evaluated at $\phi$:
\begin{equation}\label{eq:sigma}
\sigma^{b}\left[\phi,\!\Phi\right]\:\nabla_b\:\sigma^{a}\left[\phi,\!\Phi\right] \,= \,-\, \sigma^{a}\left[\phi,\!\Phi\right]\,.
\end{equation}
The defining equation \eqref{eq:sigma} can be solved by expanding in powers of $\left(\phi^a\,-\, \Phi^a\right)$:
\begin{equation}\label{eq:sigma_expanded}
    \sigma^{a}\left[\phi,\!\Phi\right] \,= \,-\,\left(\phi^a\,-\, \Phi^a\right)\,+\,\frac{1}{2}\Gamma^a_{\:\:bc}\left[\phi\right]\,(\phi^b\,-\, \Phi^b)\left(\phi^c\,-\, \Phi^c\right)\,+\,\ldots\,,
\end{equation}
where we apply the boundary conditions $\left.\sigma^{a}\left[\phi,\!\Phi\right]\right|_{\Phi\,=\,\phi}\,=\,0$ and $\nabla_b\left.\sigma^{a}\left[\phi,\!\Phi\right]\right|_{\Phi\,=\,\phi} =\,-\,\delta^a{}_b$.
After performing the saddle-point expansion, we can derive covariant expressions for the one- and two-loop parts of the 1PI effective action:
\begin{align}
   \label{eq:Gamma1loop}
   \Gamma_{1}[\varphi]\ &=\ \,-\,\frac{i}{2}\ln \operatorname{det}g_{ab}\,+\,\frac{i}{2}\ln \operatorname{det}\nabla_a \nabla_b S\ \:,\\
    \label{eq:Gamma2loop}
    \Gamma_{2}[\varphi]\ &=\ \,-\,\frac{1}{8}\Delta^{ab}\Delta^{cd}\nabla_{(a}\nabla_{b}\nabla_{c}\nabla_{d)}S\, +\: \frac{1}{12}\Delta^{a b}\Delta^{cd}\Delta^{ef}\:\nabla_{(a}\nabla_{c}\nabla_{e)}S\,\,\:\nabla_{(b}\nabla_{d}\nabla_{f)}S\:, 
\end{align}
where the round brackets $(\ldots)$ denote symmetrization over the enclosed indices. The two-loop corrections can be expressed diagrammatically as a sum of covariant Feynman diagrams
\begin{equation}
\Gamma_{2}[\varphi]\ =\ \; \raisebox{0.06\height}{\includegraphics[valign=c, scale=0.6]{images/2loops_11.pdf}}\;+\;\raisebox{0.06\height}{\includegraphics[valign=c, scale=0.6]{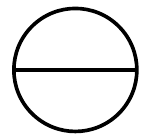}}\;.
\end{equation}

To summarize, we have built a configuration-space manifold charted by the fields $\Phi^a$ and constructed the 1PI effective action so that it remains independent of field redefinitions. Interestingly, the strength of the particles' derivative interactions is now ``felt'' by the underlying configuration space and manifests itself as geometric deformations within it.

\subsection{Geometric 2PI}\label{sec:geometric_2pi}

We now return to the 2PI formalism. To kick off this section, we redefine $K_{ab}\rightarrow K_{ab}/2$ and 
\begin{equation}\Delta^{ab}\,\rightarrow\,\Delta^{ab}\,+\,\phi^{a}\,\phi^{b}\,.
\end{equation}
(We worried a lot in previous sections about working with connected $n$-point functions and keeping track of factors of $\hbar$, but not anymore, and we will see why shortly.) With this in mind, we write the same effective action in the following more condensed form:
\begin{equation}\Gamma[\phi,\Delta]\,=\,-\,W[J,K]\,+\,J_a\,\phi^a\,+\,K_{a b}\,\Delta^{ab}\,.
\end{equation}
The discussion that follows is based on~\cite{Kluth:2023sey}, with the notation adapted for clarity and to maintain consistency with the 1PI formalism.

\begin{figure}
  \centering
   \adjustbox{scale=1.3}{
        \begin{tikzcd}[column sep=42pt,row sep=36pt]
        \Gamma[\emptyset; J, K]\hspace{2mm}
          % top horizontal (right + left)
          \arrow[r, shift left=0.6ex, yshift=-0.5ex, "J"]
          % left vertical (down K, up Δ)
          \arrow[d, shift right=0.6ex, "\Delta" swap, leftarrow]
        &
        \hspace{2mm} \Gamma[\phi; K]
          % right vertical (down K, up Δ)
          \arrow[d, shift left=0.6ex, "K"]
        \\
        \Gamma[\Delta; J]\hspace{4mm} 
          % bottom horizontal (right + left)
          \arrow[r, shift right=0.6ex, yshift=0.6ex, "\phi" swap, leftarrow]
        &
        \hspace{1mm} \Gamma[\phi, \Delta; \emptyset]
        \end{tikzcd}}
        \hspace{1.2cm}
        \adjustbox{scale=1.3}{
        \begin{tikzcd}[column sep=42pt,row sep=36pt]
        \Gamma[\emptyset; J, K]\hspace{2mm}
          % top horizontal (right + left)
          \arrow[r, shift right=0.6ex, yshift=0.7ex, "\phi" swap, leftarrow]
          % left vertical (down K, up Δ)
          \arrow[d, shift left=0.6ex, xshift=-1.5ex, "K"]
        &
        \hspace{2mm} \Gamma[\phi; K]
          % right vertical (down K, up Δ)
          \arrow[d, shift right=0.6ex, xshift=1ex, "\Delta" swap, leftarrow]
        \\
        \Gamma[\Delta; J]\hspace{4mm} 
          % bottom horizontal (right + left)
          \arrow[r, shift left=0.6ex, yshift=-0.5ex, "J"]
        &
        \hspace{1mm} \Gamma[\phi, \Delta; \emptyset]
        \end{tikzcd}}
 \caption{2PI functionals and their interrelations via Legendre transforms. The arrows indicate the variable with respect to which the Legendre transform is performed. Based on Fig.~1 of \cite{Kluth:2023sey}.}
  \label{fig:GammaDiagram}
\end{figure}
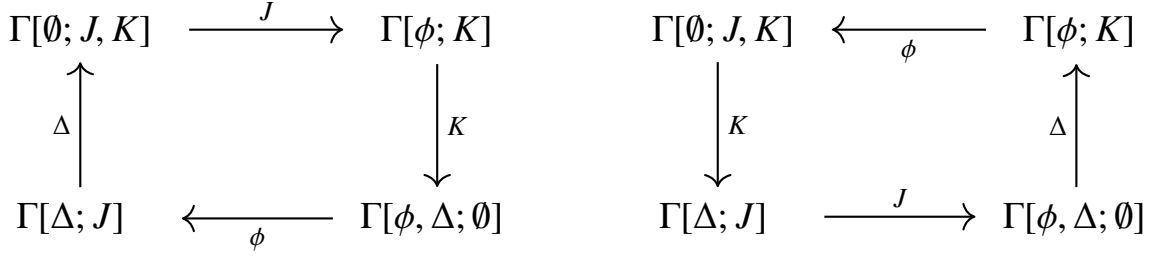

In the previous section, we noted that there exists a whole family of effective actions. In fact, even for our choice of two sources $J$ and $K$, we have four functionals, which are pairwise related by double Legendre transforms. Together with the 2PI effective action above, these four functionals are:
\begin{align}
    \Gamma[\emptyset; J,K]\,&=\,-\,W[J,K]\,,\\
    \Gamma[\phi;K]\,&=\,-\,W[J,K]\,+\,J_a\,\phi^a\,,\\
    \Gamma[\Delta;J]\,&=\,-\,W[J,K]\,+\,K_{ab}\,\Delta^{ab}\,,\\
    \Gamma[\phi,\Delta;\emptyset]\,&=\,-\,W[J,K]\,+\,J_a\,\phi^a\,+\,K_{ab}\,\Delta^{ab}\,.
\end{align}
The arguments of these functionals are arranged so that any dependence on $n$-point functions appears to the left of the semicolon, and dependence on the sources appears to the right. Note that each expression is obtained by performing a partial Legendre transform with respect to a subset of sources. The relationships among the four functionals are shown schematically in Figure \ref{fig:GammaDiagram}.

In the general $n$PI framework, we may define the effective-action functionals so that they depend on $m$ sources $J$ and $n-m$ correlation functions $\Delta$. Then, for each functional, we can define coordinates $Q$ whose upper components are the $m$ sources and whose lower components are the $n-m$ correlation functions. But let us focus on the 2PI case and the manifold arising from the Schwinger functional. We define the coordinates
\begin{equation}
    Q_\alpha\,=\,\begin{pmatrix} J_a \\ K_{ab}\end{pmatrix}\,,
\end{equation}
on the infinite-dimensional manifold of sources --- remember this is all functional --- which serves as our configuration space. The Greek index denotes both the single index associated with the one-point source and the pair of indices labeling the two-point source. 

We can now start making use of all the technology from differential geometry. Our first task is to introduce an affine connection, defined via the functional derivatives with respect to the coordinates $Q_\alpha$. Thus, given some tensor $T_{\:\:\beta_1\beta_2\dots}^{\gamma_1\gamma_2\dots}[Q]$, we can introduce the connection $D$ acting as
\begin{equation}
D^\alpha \,T_{\:\:\:\beta_1\beta_2\dots}^{\gamma_1\gamma_2\dots}[Q]\,=\,\frac{\delta}{\delta Q_{\alpha}}T_{\:\:\:\beta_1\beta_2\dots}^{\gamma_1\gamma_2\dots}[Q]\,.
\end{equation}
Note that this is just the partial functional derivative only in the $Q$ coordinate frame; we would obtain non-vanishing Christoffel symbols in other frames.
We can also define a metric based on the Hessian of some functional of the coordinates $F[Q]$:
\begin{equation}\label{eq:hessian_metric}
g^{\alpha\beta}\,=\,D^\alpha D^{\beta}F[Q]\,=\,\frac{\delta^2F[Q]}{\delta Q_{\alpha}\,\delta Q_{\beta}}\,,
\end{equation}
where the latter equality holds only in the affine coordinate frame. The geometry that results from this metric is known as \emph{Hessian geometry}, making our configuration space a \emph{Hessian manifold}, for which \eqref{eq:hessian_metric} is the defining property. Such geometries are common in information theory, so much so that we can refer to \textit{information geometry}. The Fisher metric is one example \cite{Nielsen_2020}.

The affine connection $D$ is not, in general, metric compatible (i.e., $D^\alpha g^{\beta\gamma}\,\neq\, 0$); therefore, we can introduce a tensor
\begin{equation}
    Y^\alpha_{\:\:\beta\gamma}\,=\,\frac{1}{2}D^\alpha g_{\beta\gamma}\,,
\end{equation}
named for Amari and Chentsov (see~\cite{Nielsen_2020} and references therein), which keeps track of this. We can use the Amari--Chentsov tensor to construct new connections, and two are particularly interesting. The first is the metric-compatible, Levi-Civita connection
\begin{equation}
    \nabla\,=\,D\,-\,Y \otimes \mathbb{I}\,,
\end{equation}
where the tensor product with the identity map $\mathbb{I}$ takes into account that the Amari--Chentsov tensor acts in accordance with $D$ on any given tensor~\cite{Shima}, shifting the Christoffel symbols. The second is
\begin{equation}
    \mathcal{D}\,=\,D\,-\,2Y \otimes \mathbb{I}\,,
\end{equation}
which is always affine, with affine coordinates defined as the functional derivatives of $F[Q]$:
\begin{equation}
    P^\alpha\,=\,\frac{\delta F[Q]}{\delta Q_{\alpha}}\,,
\end{equation}
such that
\begin{equation}
    \mathcal{D}_{\alpha}\,=\,\frac{\delta}{\delta P^{\alpha}}\,.
\end{equation}
Thus, we see that the coordinates $P^{\alpha}$ are dual to the coordinates $Q_{\alpha}$. Going further, we can express the inverse metric $g^{\alpha\beta}$ as the Hessian of a functional $\overline{F\,}[P]$:
\begin{equation}\label{eq:dual_metric}
    g_{\alpha\beta}\,=\,\frac{\delta \overline{F\,}[P]}{\delta P^{\alpha}\,\delta P^{\beta}}\,.
\end{equation}
The functional
\begin{equation}
    \overline{F\,}[P]=\,-\,F[Q]\,+\,Q_{\alpha}\,P^{\alpha}
\end{equation}
is nothing other than the Legendre transform of $F[Q]$. We can therefore associate $F[Q]$ with the Schwinger functional $W[J,K]~=~-~\Gamma[\emptyset;J,K]$, and $\overline{F\,}[P]$ with the 2PI effective action $\Gamma[\phi,\Delta;\emptyset]$, where
\begin{equation}
    P^\alpha\,=\,\begin{pmatrix} \phi^a \\ \Delta^{ab}\end{pmatrix}\,,
\end{equation}
and we can write
\begin{equation}
    g_{\alpha\beta}\,=\,\frac{\delta Q_{\alpha}}{\delta P^{\beta}}\,.
\end{equation}
With this technology, our $n$-point correlation functions are obtained from the action of the $D$ connection on the Schwinger functional, i.e.,
\begin{equation}\label{eq:connected_npoint}
\braket{\phi^{a_1}\ldots \phi^{a_n}}_{\rm conn.}\,=\,D^{a_1}\cdots D^{a_n}W[J,K]\,.
\end{equation}
And now we should be thankful that this connection is not metric compatible, because otherwise we would have no $n$-point functions higher than the two-point function. This can be verified by considering the defining property of the Hessian metric \eqref{eq:hessian_metric}, which, after identifying $F[Q]$ with $W[J,K]$, allows us to write \eqref{eq:connected_npoint} as
\begin{equation}
   \braket{\phi^{a_1}\ldots \phi^{a_n}}_{\rm conn.}\,=\,D^{a_1}\cdots D^{a_{n-2}}g^{a_{n-1}a_n}\,.   
\end{equation}
Therefore, if the $D$ connection were metric compatible, its action on the metric $g^{a_{n-1}a_n}$ would make any higher $n$-point function vanish.

Moving away from the affine coordinate frame, and making use of the identity
\begin{equation}
    D^\alpha \,=\, g^{\alpha\beta}\,\mathcal{D}_\beta\,,
\end{equation}
we can express everything in terms of the connected two-point function $\Delta$ and the field $\phi$, and write
\begin{equation}
\label{eq:connectedfromD}
\braket{\phi^{a_1}\cdots \phi^{a_n}}_{\rm conn.}\,=\,\left[\prod_{j\,=\,1}^{n\,-\,2}g^{a_j\,\beta_j}\,\mathcal{D}_{\beta_j}\right]\Delta_{\mathrm{c}}^{\:a_{n\,-\,1}a_n}\,,
\end{equation}
where
\begin{align}
    \label{eq:barDnonaffine}
    g^{a\beta}\mathcal{D}_{\beta}\,&=\,\frac{\delta W[J,K]}{\delta J_a\,\delta J_b}\frac{\delta}{\delta \phi^b}\,+\,\frac{\delta^2W[J,K]}{\delta J_a\,\delta K_{bc}}\frac{\delta}{\delta \Delta^{bc}}\nonumber\\
    \,&=\,\Delta_{\mathrm{c}}^{\:ab}\left[\frac{\delta}{\delta \phi^b}\,-\,\frac{\delta^2\Gamma}{\delta \phi^b\,\delta \Delta_{\mathrm{c}}^{\:cd}}\left(\frac{\delta^2\Gamma}{\delta \Delta_{\mathrm{c}}^{\:cd}\, \delta \Delta_{\mathrm{c}}^{\:ef}}\right)^{-1}\frac{\delta}{\delta \Delta_{\mathrm{c}}^{\:ef}}\right]\,,
\end{align}
with $\Delta_{\mathrm{c}}^{\:ab}=\Delta^{ab}-\phi^a\phi^b$ being the connected two-point function.

\begin{figure}
\centering
\includegraphics[width=0.6\textwidth]{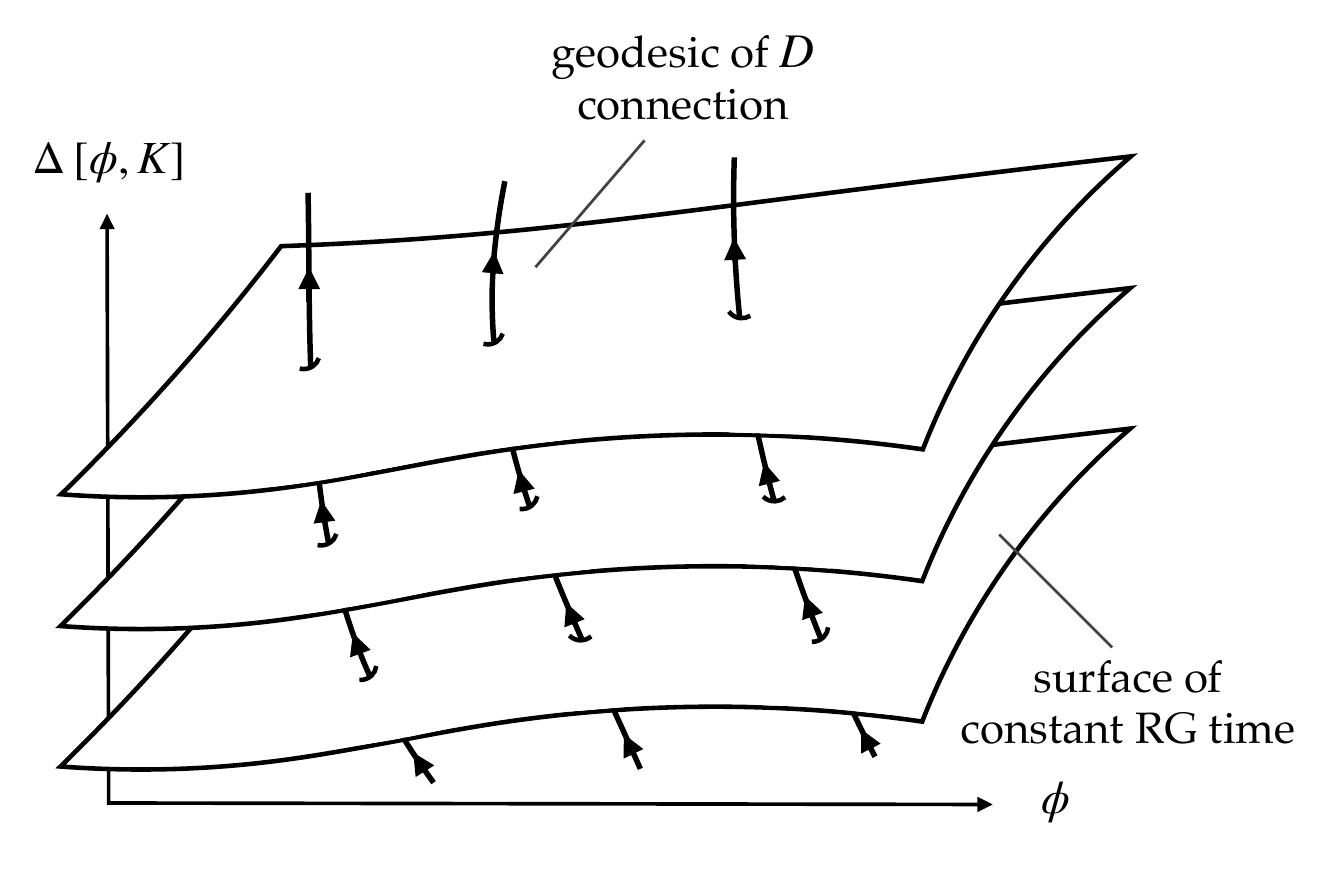}
\caption{Schematic of the affine coordinate frame of the $\mathcal{D}$ connection, spanned by $\phi^a$ and $\Delta^{ab}$. Changing the RG scale induces a vector field in configuration space, which connects hypersurfaces associated with different RG scales. All points in each hypersurface have the same value of $K_{ab}$, but different value of $\phi^a$. Geodesics of the $D$ connection are RG trajectories with $\mathrm{d}K_{ab}/\mathrm{d} t =$ constant. In the affine coordinate frame of the $D$ connection, spanned by $K_{ab}$ and $J_a$, geodesics are straight lines and pierce through hypersurfaces of constant RG time perpendicularly. For a specific example of this, see Fig.~3 of \cite{Kluth:2023sey}.}\label{fig:rg_flow}
\end{figure}

This is all very interesting, but what does it mean for the physics? Well, given that the one- and two-point functions $\phi$ and $\Delta$ are functionals of the sources $J$ and $K$, varying the sources moves us around in our configuration space or, equivalently, defines trajectories on the associated Hessian manifold. Suppose then that we let the source $K$ depend on some parameter $t$, such that
\begin{equation}
    \frac{{\rm d} K_{ab}}{{\rm d}t}\,=\,\mathrm{constant}\,.
\end{equation}
This will define a particular trajectory along which the source $K$ increases linearly with $t$. Given that
\begin{equation}\label{eq:k_2pi_relation}
    K_{ab}\,=\,\frac{\delta \Gamma}{\delta \Delta^{ab}}\,,
\end{equation}
and taking an additional derivative with respect to $t$
\begin{align}
    \frac{{\rm d}^2 K_{ab}}{{\rm d}t^2}\,&=\,\frac{{\rm d}^2}{{\rm d}t^2}\frac{\delta \Gamma}{\delta \Delta^{ab}}
\nonumber\\\,&=\,\frac{\rm d}{{\rm d}t}\left[\frac{\delta^2\Gamma}{\delta\Delta^{ab}\,\delta \phi^c}\frac{{\rm d} \phi^c}{{\rm d}t}\,+\,\frac{\delta^2\Gamma}{\delta \Delta^{ab}\,\delta\Delta^{cd}}\frac{{\rm d} \Delta^{cd}}{{\rm d}t}\right]\nonumber\\&
=\,\frac{\delta^2\Gamma}{\delta\Delta^{ab}\,\delta \phi^c}\frac{{\rm d}^2 \phi^c}{{\rm d}t^2}\,+\,\frac{\delta^2\Gamma}{\delta \Delta^{ab}\,\delta\Delta^{cd}}\frac{{\rm d}^2 \Delta^{cd}}{{\rm d}t^2}\,+\,\frac{\delta^3\Gamma}{\delta\Delta^{ab}\,\delta \phi^c\,\delta\phi^d}\frac{{\rm d} \phi^c}{{\rm d}t}\frac{{\rm d} \phi^d}{{\rm d}t}\,+\,\frac{\delta^3\Gamma}{\delta\Delta^{ab}\delta \phi^c\delta\Delta^{de}}\frac{{\rm d} \phi^c}{{\rm d}t}\frac{{\rm d} \Delta^{de}}{{\rm d}t}\nonumber\\&\phantom{=}\,+\,\frac{\delta^3\Gamma}{\delta \Delta^{ab}\,\delta\Delta^{cd}\,\delta\phi^e}\frac{{\rm d} \Delta^{cd}}{{\rm d}t}\frac{{\rm d} \phi^e}{{\rm d}t}+\frac{\delta^3\Gamma}{\delta \Delta^{ab}\,\delta\Delta^{cd}\,\delta\Delta^{ef}}\frac{{\rm d} \Delta^{cd}}{{\rm d}t}\frac{{\rm d} \Delta^{ef}}{{\rm d}t}\,=\,0\,.
\end{align}
Re-expressing the right-hand side in terms of the geometric objects defined previously (and after some relabeling of the indices), the first two terms are just
\begin{equation}
   \frac{\delta^2\Gamma}{\delta\Delta^{a_1a_2}\,\delta \phi^b}\frac{{\rm d}^2 \phi^b}{{\rm d}t^2}\,+\,\frac{\delta^2\Gamma}{\delta \Delta^{a_1a_2}\,\delta\Delta^{b_1b_2}}\frac{{\rm d}^2 \Delta^{b_1b_2}}{{\rm d}t^2} = g_{(a_1 a_2) \beta}\,\frac{{\rm d}^2P^{\beta}}{{\rm d}t^2}
\end{equation}
and the latter four terms can be written as
\begin{equation}
    \mathcal{D}_{(a_1 a_2)}\,g_{\gamma\delta}\,\frac{{\rm d}P^\gamma}{{\rm d}t}\frac{{\rm d}P^\delta}{{\rm d}t}\,=\,g_{(a_1 a_2) \beta}\,D^{\beta}g_{\gamma\delta}\,\frac{{\rm d}P^\gamma}{{\rm d}t}\frac{{\rm d}P^\delta}{{\rm d}t}\,=\,2g_{(a_1 a_2) \beta}\,Y^{\beta}{}_{\gamma\delta}\,\frac{{\rm d}P^\gamma}{{\rm d}t}\frac{{\rm d}P^\delta}{{\rm d}t}\,.
\end{equation}
Putting everything together, we arrive at the following equation:
\begin{equation}
    \label{eq:geo}
    g_{(a_1 a_2) \beta}\,\left[\frac{{\rm d}^2P^{\beta}}{{\rm d}t^2}\,+\,2Y^{\beta}{}_{\gamma\delta}\,\frac{{\rm d}P^\gamma}{{\rm d}t}\frac{{\rm d}P^\delta}{{\rm d}t}\right]\,=\,0\,.
\end{equation}
In cases where the bracket vanishes, this is nothing other than the geodesic equation of the $D$-connection
\begin{equation}
    \frac{{\rm d}^2P^{\beta}}{{\rm d}t^2}\,+\,2Y^{\beta}_{\gamma\delta}\,\frac{{\rm d}P^\gamma}{{\rm d}t}\frac{{\rm d}P^\delta}{{\rm d}t}\,=\,0\,.
\end{equation}
These trajectories are just straight lines in the affine coordinate frame of $D$, along which the $t$-derivatives of the sources are by construction conserved.  Returning to \eqref{eq:geo}, if we associate $t\propto \ln k$ with the so-called \textit{RG time}, then the geodesics of the $D$-connection on configuration space can be used to define surfaces of constant RG time or to construct RG trajectories. This scenario is depicted in Figure \ref{fig:rg_flow}.

Taking an alternative approach, we can consider the variation of the effective action with respect to $t$ while keeping $\phi$ fixed. This yields
\begin{equation}\label{eq:2pi_flow}
\partial_t\Gamma[\phi,\Delta_{\mathrm{c}}]\,=\,\frac{\delta \Gamma[\phi,\Delta_{\mathrm{c}}]}{\delta \Delta_{\mathrm{c}}^{\,ab}}\,\partial_t\Delta_{\mathrm{c}}^{\,ab}\,=\,K_{ab}\,\partial_t\Delta_{\mathrm{c}}^{\,ab}\,,
\end{equation}
where we used the fact that $\Delta$, and therefore $\Delta_\mathrm{c}$, is a functional of $K$ \footnote{ When computing subsequent derivatives with respect to $\phi^a$ and $\Delta_{ab}$, one can always exchange their order, i.e., they commute with each other. The same is not true for derivatives with respect to $\phi^a$ and $K_{ab}$, whose commutator vanishes only in special cases. More insight into this curious fact can be found in Appendix C of \cite{Kluth:2023sey}.}. This is the \textit{RG flow equation} of the 2PI effective action. 
However, to make use of~\eqref{eq:2pi_flow}, we also need to be able to obtain an expression for $\Delta_c^{ab}$ from the 2PI effective action directly.
Such an expression is conveniently provided as a result of the \textit{convexity} of the 2PI effective action \cite{Kluth:2023sey, Millington:2022cix}. Convexity is the requirement that the Legendre transform between sources and $n$-point functions be locally invertible, and, for the 2PI effective action, it reads
\begin{equation}
    \delta^\alpha{}_\beta = \frac{\delta^2 W}{\delta Q_\alpha\,\delta Q_\gamma}\frac{\delta^2 \Gamma}{\delta P^\gamma\, \delta P^\beta}\,.
\end{equation}
Making use of \eqref{eq:hessian_metric}, \eqref{eq:dual_metric} and \eqref{eq:connected_npoint}, after the identification $F\rightarrow W$ and $\overline{F\,}\rightarrow \Gamma$ and some lines of algebra, we obtain the neat relation
\begin{align}
    \Delta_{\mathrm{c}}^{\,a b}\,&=\,\left[\frac{\delta^2 \Gamma}{\delta \phi^{a}\, \delta \phi^{b}}-\frac{\delta^2 \Gamma}{\delta \phi^{a}\, \delta \Delta^{cd}}\left(\frac{\delta^2 \Gamma}{\delta \Delta^{cd}\, \delta \Delta^{ef}}\right)^{-1}\frac{\delta^2 \Gamma}{\delta \Delta^{ef}\, \delta \phi^{b} }\right]^{-1}\nonumber\\ &=\,\left[\frac{\delta^2 \Gamma}{\delta \phi^{a}\, \delta \phi^{b}}-\frac{2}{\hbar}\,\frac{\delta \Gamma}{\delta \Delta_c^{ab}}-\frac{\delta^2 \Gamma}{\delta \phi^{a}\, \delta \Delta_c^{cd}}\left(\frac{\delta^2 \Gamma}{\delta \Delta_c^{cd}\, \delta \Delta_c^{ef}}\right)^{-1}\frac{\delta^2 \Gamma}{\delta \Delta_c^{ef}\, \delta \phi^{b} }\right]^{-1}\,,
\end{align}
and this \textit{closes the flow}.\footnote{Note that the first and second lines are written in terms of $\Delta$ and $\Delta_c$ respectively, and the change of basis from $(\phi,\Delta)$ to $(\phi,\Delta_c)$ gives rise to the additional term in the second line.} Now, by acting on \eqref{eq:2pi_flow} with the $D$-connection, we can obtain systems of flow equations for the $n$-point correlation functions and, by \emph{amputating} external two-point functions, the $n$-point interaction vertices.

Thus, in the geometrical picture, we see that variations in RG time lead to changes in the correlation functions. This defines a vector field on configuration space that connects hypersurfaces of constant RG time. Therefore, the renormalization group tells us how we move through theory space, which is spanned here by the correlation functions. Table \ref{tab:geomQFT} summarizes the two geometric frameworks for quantum field theory discussed in this lecture. 

\vspace{0.5cm}
\begin{table}[h!]
\centering
\renewcommand{\arraystretch}{1.25}
\setlength{\tabcolsep}{6pt}
\begin{tabular}{p{2.5cm} p{2.7cm} p{2.5cm} p{2.5cm} p{3.6cm}}\toprule
\textbf{Framework} & \textbf{Chart\phantom{i}variables} & \textbf{Metric Type} & \textbf{Fundamental quantities} & \textbf{Example / Use} \\
\hline
Spacetime &
$x^\mu$ &
Lorentzian\:/ \hspace{5mm} pseudo-Riemannian &
$\eta^{\mu\nu}$\,,\, $R_{\mu\nu\rho\sigma}$ &
background geometry  \\
%\hline
Configuration space & 
$\Phi^a(x)$ &
pseudo-Riemannian &
$g_{ab}$\,,\, $\Gamma_{\text{1PI}}[\varphi]$ &
Vilkovisky--DeWitt \phantom{i}formalism \\
%\hline
Theory space &
$Q(J,K)$, $P(\phi, \Delta)$ &
Hessian &
$g_{\alpha \beta}$\,,\, $\Gamma_{\text{2PI}}[\varphi]$ & exact\phantom{~}renormalization group \\
\bottomrule
\end{tabular}
\caption{Compact summary of geometric frameworks for quantum field theory presented in this lecture, organized by the type of manifold, coordinate system, and key geometrical structures.}
\label{tab:geomQFT}
\end{table}

\section{Some concluding remarks}

This brings us to the end of our short quantum field theory adventure. Let us briefly summarize what we have learned so far.

We began by identifying the three fundamental ingredients that constitute a theory. In particular, we aimed to construct a quantum theory of scalar fields by means of the path-integral formalism. To this end, we first defined a classical action functional and then introduced one- and two-point external sources to probe how the system evolves away from its ground state. To generate connected correlation functions, we introduced the Schwinger functional and discovered that we can use Legendre transforms to exchange a description in terms of sources for one in terms of correlation functions. This led us to the quantum effective action, which, in the case of the 2PI effective action, generates the equations of motion for the connected one- and two-particle-irreducible correlation functions. 

While taking these steps, we were reminded of an important aspect of formulating theories: the freedom to choose different ways of charting theory space. Exercising this freedom, we also gained some geometric understanding of theory space. It turns out that the standard definition of the effective action is not inherently invariant under changes of chart variables. As the first part of our geometric excursion, we treated the fields as coordinates on an infinite-dimensional configuration-space manifold, on which we defined a pseudo-Riemannian metric and related geometric structures. We then applied this toolkit to \textit{covariantize} the 1PI effective action at one- and two-loop orders. 

For our geometric analysis of the 2PI framework, we instead took a different route and defined an infinite-dimensional manifold of sources as our configuration space. As it turns out, the most natural definition of the metric is the Hessian of some functional of the coordinates, later identified as one of the four 2PI functionals.  Exploiting the versatility of Legendre transforms, we defined a set of dual coordinates and connections, ultimately arriving at a neat physically meaningful result. By making contact with the renormalization group, we then obtained the flow equation for the 2PI effective action from the associated Hessian geometry, from which further flow equations for the correlation functions can be derived, telling us how we move around in theory space as a function of energy or length scale.

The content of this lecture is, however, merely the tip of the iceberg, and there are many topics we did not have the chance to explore. Among these are other kinds of ``integrating out,'' where one traces over a subsystem to construct an open quantum field-theoretic framework \cite{Fogedby:2022wbz}. The 2PI formalism naturally lends itself to further applications within the exact renormalization group \cite{Morris:1993qb}, which remains one of the most powerful non-perturbative techniques in quantum field theory. We didn't get to look at the role of nonlocal field redefinitions \cite{Cohen:2024fak}, quantum field theory in curved spacetime \cite{DeWitt:1975ys}, or the inclusion of fermions \cite{Gattus:2024ird, Gattus:2023gep}, gauge bosons \cite{REBHAN1987832}, or gravity \cite{Finn:2019aip}. We didn't even explore how the $\phi^4$ theory suffers from a Landau pole too and the associated issue of triviality \cite{Jafarov:2016rbw}. As it turns out, even the seemingly most innocent toy theories can prove less than upright in character. As Winnie-the-Pooh would say, ``One never can tell with theories.'' 

\section*{Acknowledgments}
This lecture is based in part on the lectures ``Advanced Quantum Field Theory,'' delivered by PM at the British Universities Summer School in Theoretical Elementary Particle Physics (BUSSTEPP), held at the University of Nottingham, 26 August -- 6 September 2025, and adapted from a series of graduate lectures delivered by PM at the University of Nottingham from 31 March -- 5 May 2016. PM would like to thank the organizers and participants of BUSSTEPP 2025 for their engagement with and feedback on these lectures. VG and PM thank Astrid Eichhorn and Yannick Kluth for helpful comments on this lecture. VG acknowledges support from the Deutsche Forschungsgemeinschaft (DFG, German Research Foundation) – SFB 1258 – 283604770. The work of PM is supported by a United Kingdom Research and Innovation (UKRI) Future Leaders Fellowship [Grant No.\ MR/V021974/2].

\end{document}